\documentclass[12pt,preprint]{aastex}

\newcommand{\GHz}{\mbox{\rm GHz} }

\newcommand{\rms}{ \mbox{\rm rms} }
\newcommand{\mkelvin}{\mbox{\rm mK} }
\newcommand{\ukelvin}{\mbox{$\mu$\rm K}}

\newcommand{\ukelvinrms}{\mbox{$\mu$\rm K rms}}
\newcommand{\ddeg}         {\mbox{${\rlap.}^\circ$}}
\newcommand{\wmap}{\mbox{\sl WMAP}}
\newcommand{\pctpyear}{\%\;\mbox{yr}^{-1}}

\newcommand{\WMAP}{\textsl{WMAP}}

\begin{document}

\title{Three-Year Wilkinson Microwave Anisotropy Probe (\wmap)\altaffilmark{1} Observations:
 Beam Profiles, Data Processing, Radiometer Characterization and Systematic Error Limits}
\slugcomment{ApJ, in press, January 5, 2007}
\author{
N. Jarosik \altaffilmark{2},
C. Barnes \altaffilmark{2},
M. R. Greason \altaffilmark{3,12},
R. S. Hill\altaffilmark{3,12},
M. R. Nolta \altaffilmark{4},
N. Odegard \altaffilmark{3,12},
J. L. Weiland\altaffilmark{3,12},
R. Bean \altaffilmark{5,13},
C. L. Bennett\altaffilmark{6},
O. Dor\'{e} \altaffilmark{5,4},
M. Halpern  \altaffilmark{7},
G. Hinshaw \altaffilmark{3},
A. Kogut \altaffilmark{3},
E. Komatsu \altaffilmark{8},
M. Limon \altaffilmark{3,12},
S. S. Meyer \altaffilmark{9},
L. Page\altaffilmark{2},
D. N. Spergel \altaffilmark{5},
G. S. Tucker \altaffilmark{10},
E. Wollack \altaffilmark{3},
E. L. Wright \altaffilmark{11}
}

\altaffiltext{1}{\wmap\ is the result of a partnership between Princeton 
                 University and NASA's Goddard Space Flight Center. Scientific 
		 guidance is provided by the \wmap\ Science Team.}
\altaffiltext{2}{Dept. of Physics, Jadwin Hall, 
Princeton University, Princeton, NJ 08544-0708}
\altaffiltext{3}{Code 665, NASA/Goddard Space Flight Center, 
Greenbelt, MD 20771}
\altaffiltext{4}{Canadian Institute for Theoretical Astrophysics, 
60 St. George St, University of Toronto, 
Toronto, ON  Canada M5S 3H8}
\altaffiltext{5}{Dept. of Astrophysical Sciences, 
Peyton Hall, Princeton University, Princeton, NJ 08544-1001}
\altaffiltext{6}{Dept. of Physics \& Astronomy, 
The Johns Hopkins University, 3400 N. Charles St., 
Baltimore, MD  21218-2686}
\altaffiltext{7}{Dept. of Physics and Astronomy, University of 
British Columbia, Vancouver, BC  Canada V6T 1Z1}
\altaffiltext{8}{Univ. of Texas, Austin, Dept. of Astronomy, 
2511 Speedway, RLM 15.306, Austin, TX 78712}
\altaffiltext{9}{Depts. of Astrophysics and Physics, KICP and EFI, 
University of Chicago, Chicago, IL 60637}
\altaffiltext{10}{Dept. of Physics, Brown University, 
182 Hope St., Providence, RI 02912-1843}
\altaffiltext{11}{UCLA Astronomy, PO Box 951562, Los Angeles, CA 90095-1562}
\altaffiltext{12}{Science Systems and Applications, Inc. (SSAI), 
10210 Greenbelt Road, Suite 600 Lanham, Maryland 20706}
\altaffiltext{13}{612 Space Sciences Building, 
Cornell University, Ithaca, NY  14853}

\email{jarosik@princeton.edu}

\begin{abstract}
The \wmap~satellite has completed 3 years of observations of the cosmic microwave background radiation. The 3-year
data products include several  sets of full sky maps of the Stokes I, Q and U parameters in 5 frequency bands,
spanning  $23$ to $94~\GHz$, and supporting items, such as beam window functions and  noise covariance matrices. 
The processing used to produce the current sky maps and supporting products represents a significant advancement
over the first year analysis, and is described herein. Improvements to the pointing reconstruction, radiometer 
gain modeling, window function determination and radiometer spectral noise parametrization are presented. A 
detailed description of the
updated data processing that produces maximum likelihood sky map estimates is presented,  along with the methods 
used to produce reduced 
resolution maps and corresponding noise covariance matrices. Finally two methods used to evaluate the noise of the 
full resolution
sky maps are presented along with several representative year-to-year null tests, demonstrating that sky maps 
produced from data
from different observational epochs are consistent. 
\end{abstract}


\section{INTRODUCTION}
 The Wilkinson Microwave Anisotropy Probe (\wmap) is a
Medium-class Explorer mission designed to produce full sky maps of the
cosmic microwave background (CMB) radiation in five frequency bands centered at
$23$, $33$, $41$, $61$ and $94$ $\GHz$. \wmap ~was launched on 2001 June 30 and began 
taking survey data on  2001 August 10 with its 20 high electron mobility transistor (HEMT)
based radiometers.  A suite of papers
\citep{bennett/etal:2003b,jarosik/etal:2003b,page/etal:2003b,barnes/etal:2003,hinshaw/etal:2003b,
bennett/etal:2003c,komatsu/etal:2003,hinshaw/etal:2003,kogut/etal:2003a,spergel/etal:2003,
verde/etal:2003,peiris/etal:2003,page/etal:2003f,bennett/etal:2003,page/etal:2003, barnes/etal:2002,jarosik/etal:2003, nolta/etal:2003}
describing the design of the observatory and the results of the first year's observations have
been previously published. Analysis of the first three years of \wmap ~data has now been completed
and the results are presented in companion papers~\citep{page/etal:prep, hinshaw/etal:prep, spergel/etal:prep}. 

One of the major design goals of \wmap~ was the careful control of systematic
errors to allow the production of high quality maps of the microwave sky.  Systematic
errors in the first year results were analyzed in several papers accompanying the data release. Analyses of
the beam shapes, beam window functions and associated errors were presented in \citet{page/etal:2003b}.
Details of the data processing and related errors were discussed in \citet{hinshaw/etal:2003b}.
Radiometer performance and systematic errors were analyzed in \citet{jarosik/etal:2003b} and 
\citet{hinshaw/etal:2003b} while errors related to sidelobe pickup were 
explored in \citet{barnes/etal:2003}.

This paper updates and extends the previous analyses through the use of three years of observational data
and addresses additional issues, such as data processing specific to the production of polarization maps.
The overall instrument performance and 
improvements to the instrument modeling are described in \S~\ref{sec:inst_perf_and_model}.
 Included in this section are updates to the radiometer gain models and
beam window function analysis. In \S~\ref{sec:data_proc} changes and additions to the data
processing procedures, including the technique used to produce the maximum likelihood
sky maps for Stokes I, Q and U parameters,  and the evaluation of the pixel-pixel inverse 
noise matrix are described. In \S~\ref{sec:tests_on_data} tests
on the sky maps, including year to year comparisons and evaluation of map noise levels, are discussed.

\section{INSTRUMENT PERFORMANCE AND MODELING}\label{sec:inst_perf_and_model}
\subsection{Observatory Status and Observing Efficiency}
The \wmap \ spacecraft and instrument 
continued to perform normally throughout its second and third years
of science data collection, spanning 2002 August 10 through 2004 August 9.
Two station-keeping maneuvers were performed during each of the second and third years of \wmap\ observations.
Each of these maneuvers resulted in the loss of approximately 10 h
of science data, approximately 0.5 h for the maneuver itself and the remainder for the recovery from
the thermal disturbance to the observatory. The observatory also 
entered a safe-hold mode on 2003 10 Aug,  believed to 
be the result of a cosmic ray event. This occurrence resulted in a loss of approximately 60 h of data, caused both by the
time spent in safe hold mode itself, and the time required for observatory temperatures to re-stabilize after 
the associated thermal disturbance. These 
periods, plus a short ($\approx 3$ h)
data loss due to a data recorder  overflow caused by ground station difficulties that occurred on 2003 May 25, resulted in the loss 
of a total of 103 h of science data,
resulting in an overall observing efficiency of $99.4~\%$ for \wmap's second and third years of observations.
A tabulation of the times of the data excluded from processing by the aforementioned events
can be found in \citet{limon/etal:prep}. 

During the second and third years of operation, 8 sudden jumps were observed in the outputs of the
radiometers. 
These events, distributed among 5 of the 20 radiometers  comprising \wmap, are believed to be 
the result of sudden releases of thermally induced  mechanical stresses that slightly alter
the properties of some radiometer components \citep{jarosik/etal:2003b}. During the first
year of operation, 21 such events were observed. It is believed that the reduction in the number
of events is the result of a more stable thermal environment during the second and third years of operation.
Although the events are of short duration, the data processing used to produce the sky maps
requires that a 2-4 h segment of data for the affected radiometer be excluded from the 
final processing to avoid
artifacts in the sky maps. All 8 events taken together caused a loss of $0.015~\%$ of the science
data from the second and third year of observations. 

\subsection{Instrument Thermal Environment}
\label{sec:thermal}
Providing a stable thermal environment for the \wmap\ instrument is a key component in the production of a
high quality data set with well defined systematic error limits. Of particular importance are temperature
variations synchronous with the 129 s spin period of the observatory. The methods used
to design and predict the thermal environment of the instrument and how they affect the
instrument performance are described in 
\citet{bennett/etal:2003} and \citet{jarosik/etal:2003}. On orbit, 56 high-resolution platinum-resistance
thermometers monitor the temperature of key instrument components. These data provide verification that
the actual thermal stability meets design specifications and allow for estimates of spurious
signal levels resulting from radiometer thermal fluctuations. Analysis of the flight thermal data 
indicates the expected  annual temperature
modulation arising from the eccentricity of \wmap's orbit and the asymptotic gradual warming 
of the instrument components as the thermal blankets degrade from 
ultraviolet exposure. Values characterizing the annual 
temperature modulation and gradual warming of the
instrument components are presented in Table~\ref{table:thermal_perf}. The slow warming of the 
instrument components has increased the radiometer noise levels 
by $< 0.1~\pctpyear$ based on analysis of the noise in the sky maps as presented in \S~\ref{sec:map_noise}.
 Projections indicate that the \wmap~ observatory should be able to operate
for years (with rare solar eclipse avoidance maneuvers). Plots of the instrument and observatory temperature
histories are presented in \citet{limon/etal:prep}.

\begin{figure*}
\epsscale{0.7}
\plotone{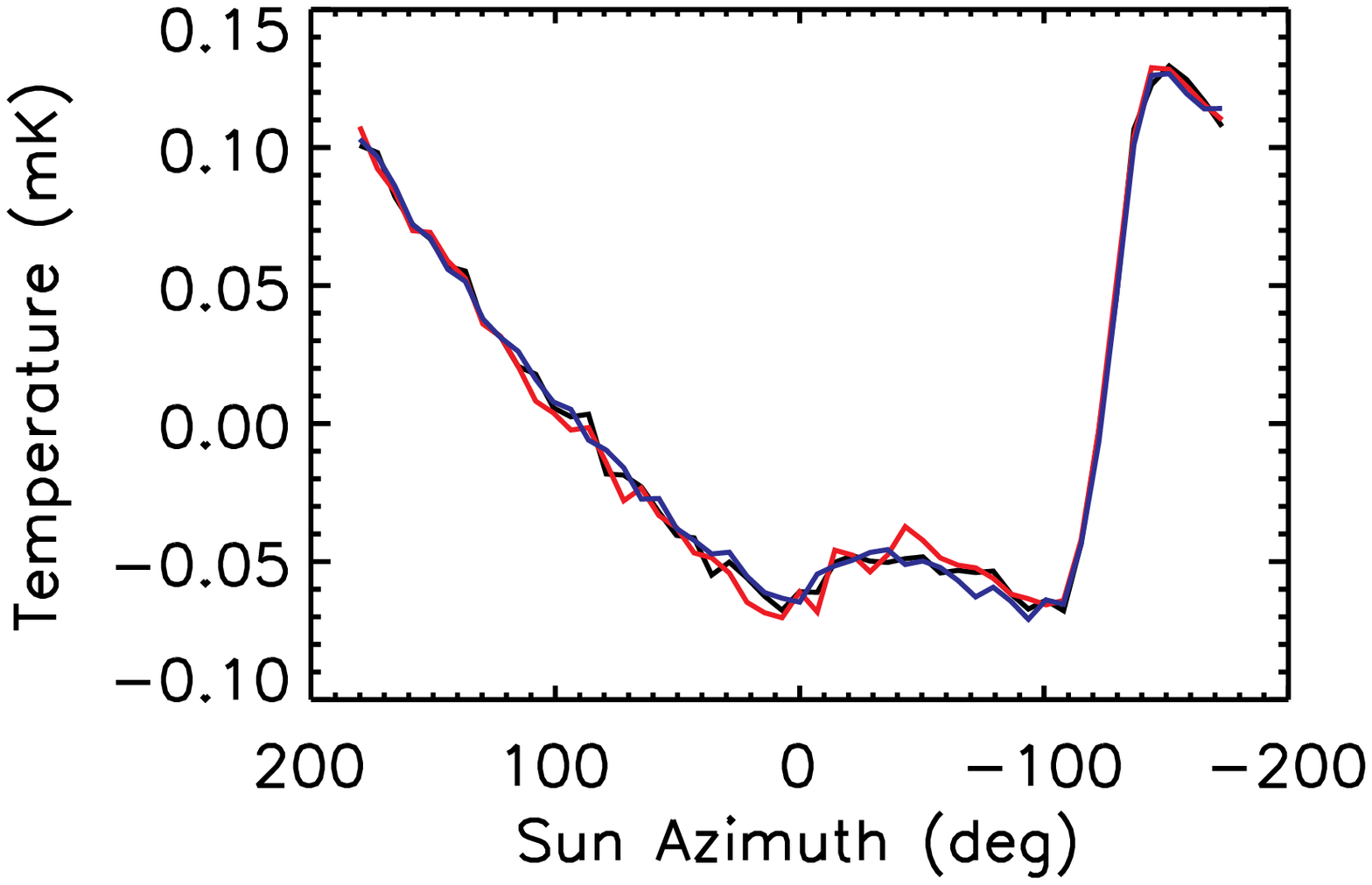}
\caption{The temperature of the middle of the B side primary mirror vs. the Sun azimuth in the spacecraft 
coordinate system. The red data were taken during the first year of observations, the black during
the second and the blue during the third. The similarity between the three 
indicates that the screening of the cold optics from
solar radiation is still effective during the second and third years of operation. }
\label{fig:DTBMIDPRIT}
\end{figure*}

Significant spin-synchronous temperature variations are detected in 10 of the
13 thermometers attached to the thermal reflector system (TRS) with peak-to-peak
variations ranging from $15 - 198~\ukelvinrms $. Similar variations were seen
during the first year of observations \citep{page/etal:2003b, limon/etal:2003} and were identified
as the result of impulsive heating by solar radiation scattered off  the rim of the
solar array.  Figure~\ref{fig:DTBMIDPRIT} displays the largest of these signals
for each year of observations. The measured temperature profile as 
a function of Solar 
azimuth with respect to the observatory follows the same pattern for all 3 years of data. 
 The similarity of
the temperature profiles indicates that no significant degradation has occurred in the solar
screening properties of the solar arrays or insulation blankets. 
The predicted microwave emissivity
of the reflector surface ($0.0005$) yields an estimated spin synchronous peak-to-peak 
spurious signal of $\approx 0.2~\ukelvin$ in the time ordered data (TOD). 
   
\begin{deluxetable}{cccclc} 
\tablecaption{\wmap~Year-2 and Year-3 Thermal Characteristics\label{table:thermal_perf} }
\tablecomments{ This Table summarizes the thermal environment of the \wmap~
observatory during its second and third years of observations. The ranges specified for 
the ``Mean'' are the 24 month averages 
of the hottest and coldest thermometer in each assembly. The average increase in temperature
year-3 relative to year-2 is given as $\Delta T_{\rm avg}$. ``Annual Mod'' is the amplitude of the
annual temperature modulation arising from the eccentricity of the \WMAP~orbit. 
The measured spin synchronous (Meas. SS.) values are obtained by
binning measured temperatures  in  spin synchronous spacecraft coordinates relative
to the Sun, after spline removal of long term temperature drifts. The values presented are
derived from one month of data and are the largest limits obtained for any thermometer 
within each assembly. No significant spin synchronous temperature variations are observed for the
Power Distribution Unit, Analog Electronics Unit, Receiver Box or Focal Plane Assembly, 
so the limits presented (indicated by $<$ ) are determined by the 
binned readout noise of the thermometers. Spin synchronous temperature variations are detected in
the Thermal Reflector System  and are at levels similar to those observed during the first year
of operation. The last column contains the design requirement for the spin synchronous temperature stability of each 
assembly. \emph{The measured spin synchronous temperature fluctuations are all far below the maximum design 
values.}}
\tablecolumns{6}
\tablehead{
\colhead{Component}&
\colhead{Mean}&
\colhead{$\Delta T_{\rm avg}$}&
\colhead{Annual Mod}&
\colhead{Meas. SS.}  &
\colhead{Req. SS.}  \\ 
 & (K) & (K) &(K) &   (\ukelvinrms) &  (\ukelvinrms)} 
\startdata
Power Distribution Unit    & 297.4-300.0 &0.98 &0.97  & $< 6.9$ & $ < 10000$ \\ 
Analog Electronics Unit    & 300.3-307.4 &0.87 &0.80  & $< 8.3$ & $ < 10000$ \\
Receiver Box               & 287.1-289.2 &0.44 &0.54  & $< 8.0$ & $ < 500$   \\
Focal Plane Assembly       & 88.2-91.6   &0.19 &0.16  & $< 5.7$ & $ < 500$   \\
Thermal Reflector Assembly & 67.6-87.1   &0.16 &0.16  & $\sim 200$ & $ < 5000$  \\     
\enddata
\label{tab:thermal_stab}
\end{deluxetable}
   
No spin-synchronous temperature variations were detected in the
Focal Plane Assembly (FPA), which contains the cold radiometer components, the
Receiver Box (RXB), which contains the warm radiometer components and phase switch drivers,
the Analog Electronics Unit (AEU) containing the data collection electronics, or 
the Power Distribution Unit (PDU) which contains  the power supplies for the HEMT amplifiers. 
The limits for such signals are presented in  Table~\ref{tab:thermal_stab}. The bounds
are limited by thermometer readout noise and are similar to the year-1 values.  
Limits on possible spurious
signals arising from thermal fluctuations of the radiometers are obtained by 
combining the on-orbit temperature variations measured during the second and 
third years of \wmap~observations with instrument thermal 
susceptibility coefficients based on preflight measurements \citep{jarosik/etal:2003b} 
and on-orbit  measurements \citep{hinshaw/etal:2003b}.  Artifacts in the  
TOD resulting from thermal fluctuations
of radiometer components are found to be less than $0.4~\ukelvinrms$ for all 20 radiometers, consistent
with the year-1 results. \emph{Given the low level of the expected spin synchronous signal, no corrections
for spin synchronous effects have been applied  to the time ordered data in the 3-year  analysis.}

\subsection{Pointing Determination} 
\label{sec:pointing}
The boresight directions of the \wmap\ beams
relative to the observatory are measured by relating radiometric measurements
of Jupiter to the attitude of the observatory as determined by two onboard
star trackers \citep{hinshaw/etal:2003b}. \wmap\ observes Jupiter
 $\approx 90$  days per year in two $\approx 45$ day ``Jupiter seasons''. To test pointing stability, boresight directions are
computed separately for each Jupiter season, and differences between the
individual seasonal determinations are noted.  During the first year of \wmap\ observations,
the azimuthal beam positions found for the first two Jupiter seasons agreed to better than
$ 3^{\prime\prime}$, but the elevation positions differed by $\approx 10^{\prime\prime}$.  
This small difference was
consistent with expected error in the spacecraft quaternions and was treated as
part of the error budget rather than being actively corrected.
 
The second year of \wmap\ operation provided two additional Jupiter observing seasons,
2002 Nov 01 - 2002 Dec 26 and 2003 Mar 13 - 2003 May 08. With the addition of
the third season, it was found that, while azimuthal positions were stable,
the apparent elevations of the beams relative
to the spin axis of the satellite now differed by $ \approx 30^{\prime\prime}$ from the positions originally
computed from the first Jupiter season. Jupiter observations from season 4 confirmed the
systematic change in elevation.  By comparing the attitude measurements of the two
star trackers and the radiometric Jupiter observations, the systematic variation was
traced to a temperature dependent flexure occurring in the spacecraft structure, which
altered the pointing of the star trackers relative to the microwave telescope's beams.
 
Processing of the 3-year \wmap~ data assumes a linear
dependence of apparent star tracker elevation with spacecraft temperature, and
corrects the spacecraft quaternions appropriately. 
A new set of line-of-sight vectors
describing the telescope beam positions for use with the updated quaternions
is available with the data release\footnote{ All data products in the current release are 
available through \texttt{http://lambda.gsfc.nasa.gov}}.
The resulting pointing corrections are small compared to the size of the beams $(\geq 12^\prime)$ and therefore
produce negligible changes to signals in the maps, but will slightly change the noise
patterns since observations near pixel borders may have moved into adjacent pixels relative to the first year analysis.
Based on the data from the six Jupiter seasons, residual pointing errors after
application of the corrections are estimated to be $ < 10^{\prime\prime}$. A detailed description of the 
modeling of the pointing correction  is contained in~\citet{limon/etal:prep}.
 
\subsection{Radiometer Gain Model} 
\label{sec:gain_model}
The year-1 analysis of the \wmap~ data employed a model that related small 
changes in the radiometer gains (typically $\approx$ 1 \%) to values of radiometer housekeeping data
based on a physical model of the radiometer performance \citep{jarosik/etal:2003b}. Each detector was
modeled separately and used the measured values of the FPA temperature, $T_{\rm FPA}$, 
the radio frequency (RF) bias powers on the detectors, $\overline{V}$, and three parameters determined by fitting
the model to the hourly gain measurements, obtained from the CMB dipole signal.
The form of the model used in the year-1 analysis was
\begin{equation}
G = \alpha \frac{\overline{V}-V_0}{T_{\rm FPA}-T_0}. \label{eqn:p1_gain_model}
\end{equation} 
 The parameter
$V_{\rm 0}$ characterizes compression (non-linear response) of the amplifiers and detectors, $T_{\rm 0}$ 
characterizes the variation of the  noise temperature of the input amplifiers  with
physical temperature, and $\alpha$  normalizes the overall gain. An example of the
gain model performance fit to the hourly dipole-based gain measurements can be found in
\citet{jarosik/etal:2003b}.

\begin{figure*}
\epsscale{0.8}
\plotone{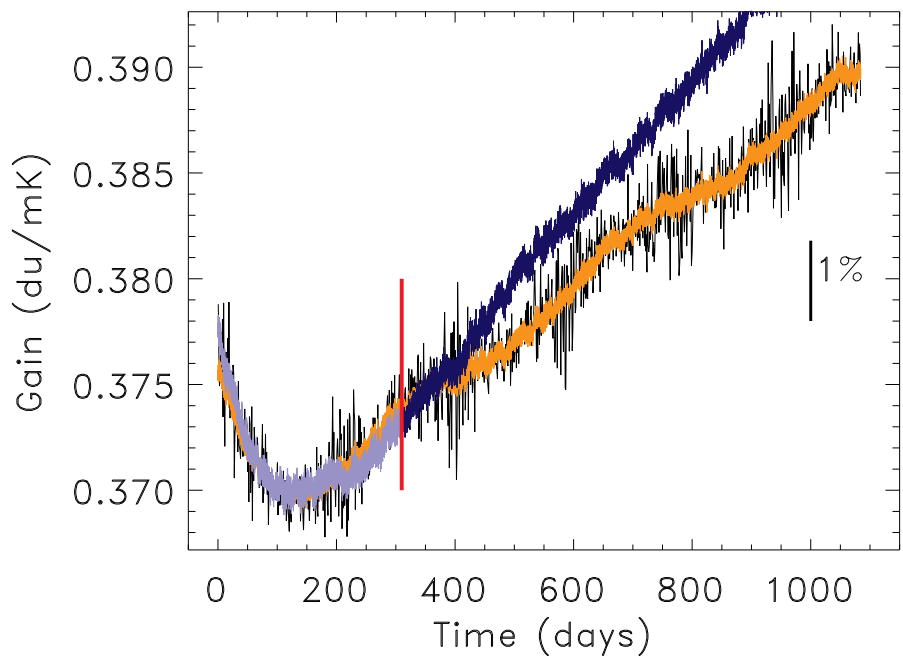}
\caption{Comparison of the hourly gain determinations (black) based on measurement of the CMB dipole to
two different versions of the radiometer gain model. These data are for the V223 detector 
and the time range spans the three years of \wmap~science data collection. 
The blue lines are the original gain model  derived by fitting the initial 310 days of data. The light blue region, 
to the left of the vertical red line, indicates the time range used to fit the model.
The dark blue region, to the right of the red vertical line, is this model as originally fit, 
applied to the  remainder of the data. The orange line is 
the new form gain model fit to all three years of data. The updated gain model is a significantly better fit to the hourly
gain measurements. Note that the difference between the models contains a component roughly linear in time.}
\label{fig:gain_model} 
\end{figure*}

When the gain models fit to the first year observation data were extended to the 
3-year data significant deviations became evident. These deviations became apparent
as the input housekeeping data to the model ($\overline{V}$ and $T_{\rm FPA}$) spanned 
a larger range than was used to fit the model, due  to the slow warming of the 
observatory. Re-fitting the original gain model parameters ($V_{\rm 0}, T_{\rm 0}$ and
$\alpha$) greatly improved the agreement between the model and dipole derived gain measurements,
but significant errors still were evident. An additional 
term was added to the model to account for
the \emph{variation} of the compression of the amplifiers and 
square law diode detectors with their physical temperature by allowing 
the parameter characterizing the nonlinearity to have a weak temperature dependence. 
The resulting form of the gain  model is
\begin{equation}
G = \alpha \frac{\overline{V}-V_0 -\beta(T_{\rm RXB} - 290)}{T_{\rm FPA}-T_0},\label{eqn:p2_gain_model}
\end{equation} 
where $T_{\rm RXB}$ is the temperature (in Kelvin) of the RXB thermometer  closest to the
detector and $\beta$ is an additional parameter to be fit. Figure~\ref{fig:gain_model} shows the dipole
derived gain measurements for the V223 detector and the improved gain model for this detector,
fit to the entire 3 years of data. Also shown in this Figure is the original gain model, using model parameters fit to year-1 data only, 
but applied to the entire 3 years of data . The new model significantly improves the fit 
over the entire 3 year period. The first-year analysis implemented a small, time dependent weighting of 
the TOD using the denominator of equation~(\ref{eqn:p1_gain_model}) as a measure of the input referenced 
noise temperature of the HEMT amplifiers. The values of $T_0$ and $\beta$ are strongly coupled in the current gain model and
are not independently determined with high accuracy. 
It is therefore not possible to use the denominator of equation~(\ref{eqn:p2_gain_model}) as a measure of the radiometer
 noise levels. The three year processing therefore assumes uniform radiometer noise within
each year. We maintain the year-1 estimate of the absolute sky map calibration uncertainty of  $0.5~\%$. 

\subsubsection{ The Gain Model's Effect on Measurement of the Quadrupole}\label{sec:gain_model_quad}
Comparison of the year-1 and 3-year gain models indicated that many of the year-1 gain models displayed
small errors in the predicted gain roughly linear in time,  with an error on the order of 
$0.3~\pctpyear$. 
\begin{figure*}
\epsscale{0.8}
\plotone{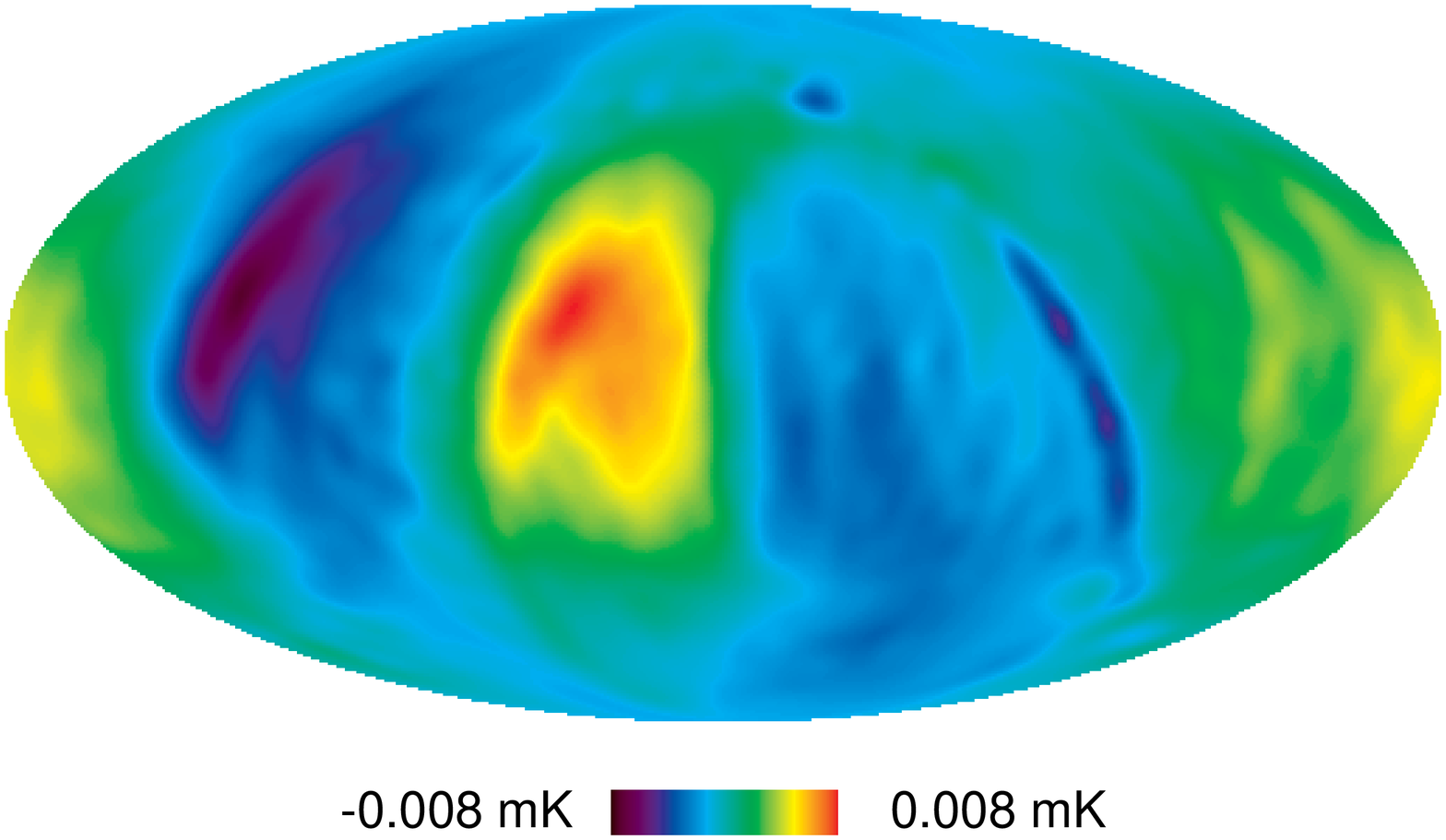}
\caption{ Difference between the temperature sky maps produced using two different 
gain models. Raw data from from the V2 differencing assembly for year-1
was processed both with the original (year-1) and the improved (3-year) gain models.  
The map projection shown is in ecliptic rather than Galactic coordinates.  The
observed quadrupolar feature
arises from imperfect subtraction of the velocity induced dipole signal in maps processed with
the year-1 gain model. Similar features are observed in similarly constructed difference maps for
 many of the DAs.}
\label{fig:v2yr1-v2yr2}
\end{figure*}
\begin{figure*}
\epsscale{0.8}
\plotone{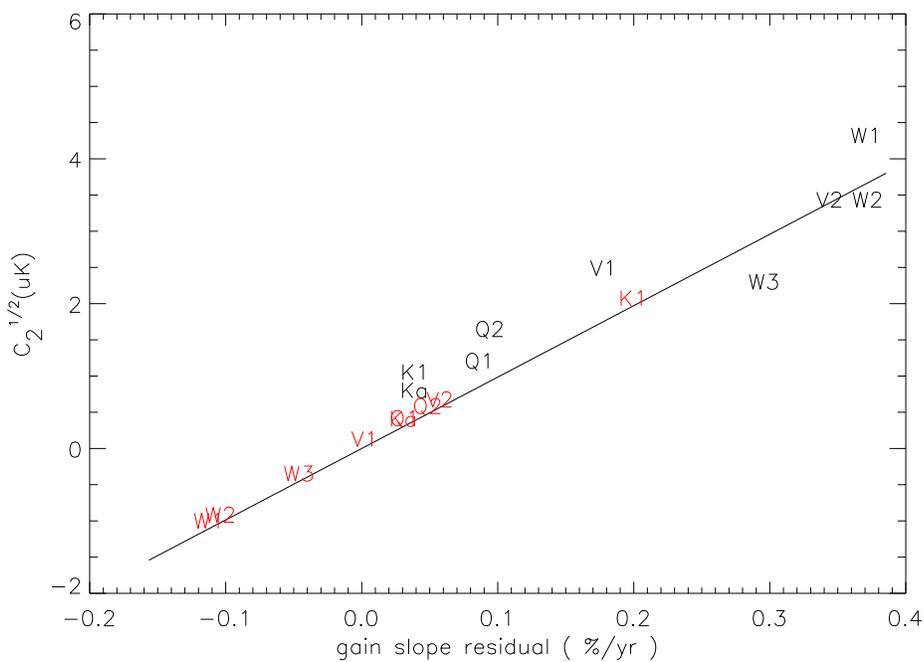}
\caption{ Amplitude of the quadrupole difference between temperature maps produced using 
the original and improved gain models. The horizontal axis is the difference between the
mean slopes of the gain models.
Each black symbol represents one Differencing Assembly. The
line is a fit to these points constrained to pass through the origin and has a 
slope of  $9.9~\ukelvin/(\pctpyear)$. The red symbols are estimates
of the amplitude of residual quadrupolar artifacts in the temperature maps after application of the
improved gain model. Note that these values are distributed around zero indicating that the phases of the
estimated residuals are random and should not bias the measured sky quadrupole.}
\label{fig:c2_vs_dg}
\end{figure*}
For each DA\footnote{ A Differencing Assembly (DA) is a pair of differential radiometers connected to
the two linear polarizations of a set of telescope feed horns.}, difference 
maps were formed by subtracting maps  processed with
the original gain model from those processed with the
improved gain model. These maps displayed  a feature with a several micro-Kelvin quadrupolar component, an example of
which is presented in Figure \ref{fig:v2yr1-v2yr2}. 
This signal was found to have similar morphologies in many of the maps with its 
amplitude correlated to the difference between the mean slopes  
of original and improved gain models. 
No other multipoles show significant correlation with the mean slopes of the gain models. 
(Sinusoidal differences between the gain models with an annual period were also 
examined and showed no correlations with any multipole of the difference maps.)
These small errors in the gain model resulted in errors in subtraction of the  $\approx 3~\mkelvin$ 
dipole signal.
When processed through the map-making pipeline this residual dipole signal yielded a 
systematic feature with a  quadrupolar component  common to  
many of the sky maps. 
Since the form of the residual gain error was common to many of the radiometers the 
resultant spurious quadrupolar features were aligned, resulting in a biased
measurement of the quadrupole moments in both the auto and cross power
spectra \citep{hinshaw/etal:2003}. The alignment of this spurious feature was such that it  partially canceled
the true sky quadrupole signal. Correction of this systematic error accounts for a part of the small increase in the 
reported value of the quadrupole moment, $\ell(\ell+1)C_2/2\pi$,  from $154~\ukelvin^2$ in the year-1 
release~\citep{bennett/etal:2003b} to $220~\ukelvin^2$, the largest change being the result of the change of the estimator use
to evaluate the quadrupole ~\citep{hinshaw/etal:prep}. It should be noted that the current value is still far smaller than the mean expected value of 
$\sim 1220 \ukelvin^2$ obtained from the best fit $\Lambda$-dominated cold dark matter models~\citep{hinshaw/etal:prep}.

Estimates of residual quadrupolar contamination remaining in the maps processed with 
the improved gain model  are obtained 
by multiplying a coefficient relating the amplitude of the spurious quadrupolar signal to the slope
error in the gain model, $\delta{c_2}/ \delta{\dot{g}}$,  by an estimate of the residual slope error in 
the improved gain model, $\Delta{\dot{g}}$.
The value of    $\delta {c_2}/\delta{\dot{g}} = 9.9~\ukelvin/(\pctpyear)$ was 
obtained by a linear fit
to the data contained in Figure \ref{fig:c2_vs_dg} constrained to pass through the origin. 
The residual slope error in 
the gain model, $\Delta \dot{g}$, was estimated as the sum of two terms: a fit to the residuals of the gain model
to the dipole based gain measurements, and a term to allow for a possible bias in the slope
of the dipole based gain determinations. The value of this second term was obtained from simulations of the complete 
calibration algorithm including effects of far sidelobe pickup of Galactic and dipole signals. These simulations 
indicate that the hourly dipole based gain measurements have slope biases ranging from $+0.03 $ to $ -0.04~\pctpyear$. 
The magnitudes of these biases were summed with the magnitudes of the residual errors between the
dipole based gain determinations and the gain model to yield an estimate  $\Delta \dot{g}$.  Values 
of  $\Delta \dot{g}$ are all less than $0.23~\pctpyear$.
The resulting estimates of the magnitude of possible spurious quadrupolar signals in the 3-year maps
are plotted in red in Figure \ref{fig:c2_vs_dg} . 
An estimate of spurious  contributions to $C_2$ is obtained by calculating the
quantity
\begin{equation}
\Delta C_2 = 2 \left<\frac{\delta{c_2}}{\delta{\dot{g}}}\ \Delta \dot{g}\right> C_2^{\rm nom} 
\end{equation}
where the average is over the Ka-W band DAs, and the nominal value for the quadrupole moment is taken as
 $\ell(\ell +1) C_2^{\rm nom}/2\pi = 236~\ukelvin^2$. The resultant estimate of the uncertainty on the quadrupole moment arising from
gain model errors, $\Delta C_2$,  is $31~\ukelvin^2$. 

\subsection{Time Ordered Data Power Spectra} \label{sec:TOD_power_spectra}
\begin{figure*}
\epsscale{0.7}
\plotone{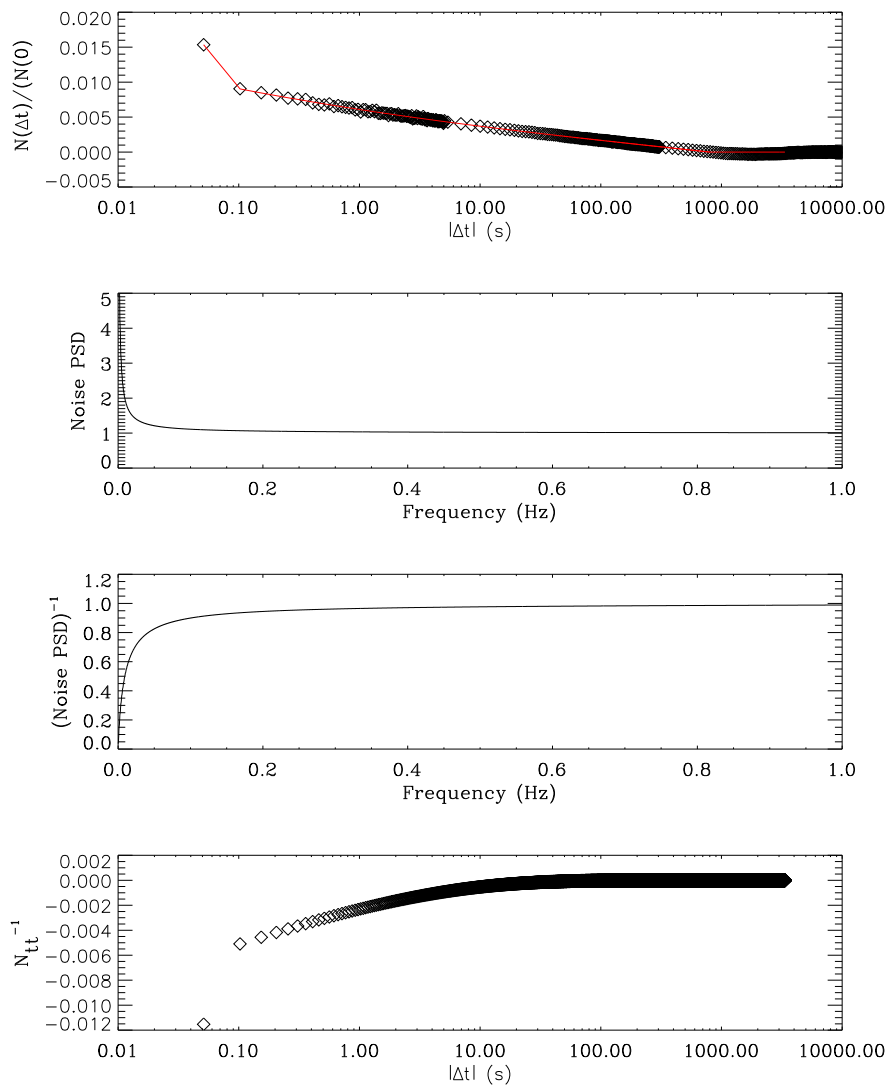}
\caption{Noise and Filter Properties of the W11 Radiometer. The top panel displays the measured autocorrelation 
function of the W11 radiometer noise 
(black diamonds) and the parameterized fit to these data (red line) as described in \S~\ref{sec:TOD_power_spectra}. The data point
at $\Delta t= 0$ with value 1 has been omitted for clarity. The remaining plots illustrate the steps used to form the $N_{\rm tt}^{-1}$
filters used in the conjugate gradient map solution, described in 
\S~\ref{sec:map_evaluation}. The second panel displays the noise power spectral density obtained from a
Fourier transform of the parameterized noise autocorrelation function, while the third panel shows the reciprocal 
of this function. The last panel presents
the $N_{\rm tt}^{-1}$ filter function obtained via a Fourier transform of the reciprocal noise power spectral density, again with the 
data point at $\Delta t = 0$ and value 1 omitted for clarity.}
\label{fig:ACF_filter}
\end{figure*}

The outputs of the \wmap~radiometers exhibit excess power at low frequencies
characterized by a ``1/f knee frequency'', $f_{\rm knee}$ as described in 
\citet{jarosik/etal:2003b}. Production of sky maps involves application of a
filter to the TOD, as described in 
\S \ref{sec:cgmaps}.
These filters are derived from fits to the autocorrelation
function of the TOD after removal of an estimated sky signal based on preliminary sky maps. 
The measured autocorrelation functions
for each radiometer are parameterized as
\begin{equation}
N(\Delta t) = \left\{ \begin{array}{lll}
AC, & \mbox{$\Delta t = 1$,} \\
a + b \log(|\Delta t|) + c [\log(|\Delta t|)]^2 + d [\log(|\Delta t|)]^3,  & \mbox{$ 1 < |\Delta t| < \Delta t_{\rm max}$,}\\
0 & \mbox{$ |\Delta t| \geq \Delta t_{\rm max}$}
\end{array}
\right. 
\end{equation}
where $\Delta t$ is the time lag between data points in units of samples, the parameters $AC$, $a$, $b$, $c$ and $d$
are determined by fitting to the autocorrelation data, and $\Delta t_{\rm max}$ is the time lag at which the fit
crosses zero, typically $\approx 600$ s. 
These fits were performed on a year-by-year basis to allow for gradual changes in the radiometer noise characteristics, 
even though the  radiometer noise properties are very stable. An example of this procedure is presented in the top panel 
of Figure \ref{fig:ACF_filter}, which presents the measured autocorrelation data and the parameterized fit for
the year-3 observations of the W11 radiometer.
Tabulated values of the parameterized filter coefficients for all years and DAs are available with the data release.

\subsection{ Transmission Imbalance Measurement}
The year-1 \wmap~analysis included a measurement of a set of transmission imbalance factors, $x_{\rm im}$, characterizing 
the different transmission of sky signals from the A side and B side optics into the radiometers.
The response, ${\bf d}$, of the \wmap~radiometers to sky signals $T_{\rm A}$ and $T_{\rm B}$ may be written
\begin{equation}
{\bf d} \propto (1 + x_{\rm im}) T_{\rm A} -  (1 - x_{\rm im}) T_{\rm B}, \label{eqn:loss_imbalance_def}
\end{equation} 
where the $x_{\rm im}$ factors parameterize a departure from ideal differential radiometer performance,
specifically the level of response to common mode input signals~\citep{jarosik/etal:2003b}. For an ideal differential 
radiometer $x_{\rm im} = 0$,  and the radiometer exhibits no response to common mode signal.

The transmission imbalance factors are determined by fitting the raw TOD to templates composed
of pure differential and pure common mode signals originating from the CMB dipole anisotropy~\citep{jarosik/etal:2003b}.
The year-1 analysis was performed using 232 days of data, and did not remove an estimate of Galactic
emission and CMB anisotropy before performing the fits, since reliable sky maps were not 
available at the time the analysis was performed. The 3-year analysis improves on the previous analysis in 
two respects; 1) estimates of the Galactic and CMB anisotropy signals are removed from the TOD, leaving only the dipole signal, 
 before fitting to the aforementioned templates, and 2) three years of TOD are used in the analysis. 
A fit to all three years of data was used to determine the values of the transmission
imbalance factors, which are presented in Table~\ref{tab:transmission_imbalance}. These values agree with those
from the year-1 analysis to the degree expected, but are more accurate due to the improvements in the 
processing.

\begin{deluxetable}{lllll}
\tablecaption{ Input Transmission Imbalance Measurements of the \wmap~Radiometers\label{tab:transmission_imbalance}}
\tablehead{
\colhead {Radiometer} & \colhead{$x_{\rm im}$} & \colhead{\phantom{xxxxx}} &
\colhead{Radiometer} &\colhead{$x_{\rm im}$}} 
\startdata
       K11&       $0.0000 \pm 0.0007$ &      & K12&       $0.0056 \pm 0.0001$\\
      Ka11&       $0.0035 \pm 0.0002$&      &Ka12&       $0.0014 \pm 0.0002$\\
       Q11&       $0.0010 \pm 0.0003$&      & Q12&       $0.0047 \pm 0.0006$\\
       Q21&       $0.0075 \pm 0.0012$&      & Q22&       $0.0103 \pm 0.0007$\\
       V11&       $0.0011 \pm 0.0003$&      & V12&       $0.0027 \pm 0.0003$\\
       V21&       $0.0043 \pm 0.0006$&      & V22&       $0.0051 \pm 0.0021$\\
       W11&       $0.0092 \pm 0.0024$&      & W12&       $0.0023 \pm 0.0010$\\
       W21&       $0.0103 \pm 0.0019$&      & W22&       $0.0084 \pm 0.0018$\\
       W31&       $0.0014 \pm 0.0015$&      & W32&       $0.0045 \pm 0.0010$\\
       W41&       $0.0208 \pm 0.0034$&      & W42&       $0.0219 \pm 0.0062$
\enddata
\tablecomments{ 
Measurement of the fractional input transmission imbalance,  $x_{\rm im}$,
obtained from the \wmap~3-year data. They were obtained
via measurements of the radiometer responses to the common mode signal arising from the CMB dipole.
All the values are small, nevertheless corrections for this effect have been included in the
map making algorithm. The values of the uncertainties are estimated from the variance of  $x_{\rm im}$
measurements from 3 single year analyses.}
\end{deluxetable}

\subsection{Beam and Window Function Determination}\label{sec:beams_and_windows}
Knowledge of the optical beam shapes is of critical importance to the 
scientific interpretation of the data. The shape of the measured 
power spectrum, and through that the values of the cosmological parameters,
depends on the beam window functions.
In turn, the beam window functions are determined from  
the shapes of the beam profiles. Uncertainties in the beam profile at the
-20~dBi level (30 to 40 dB down from the peak, near the noise level of 
the measurements) can influence the window function at the 1\% level. 
This sensitivity has motivated an extensive investigation of the beams. 

The beam modeling we present here is based on six seasons of Jupiter 
data acquired over three years of observations. The modeling has been
significantly enhanced and the approximations made for the first year 
analysis have been reexamined. 
With the new models, the beam profiles can be extrapolated
below the noise level of the maps, thereby obviating
the cutoff radius, $\theta_{\rm Rc}$, introduced in \citet{page/etal:2003}.
It is found that on average the beam solid angles are systematically 
larger by 1\% than the year-1 estimates, leading to a systematic decrease
in the window functions at $\ell>50$, leading in turn to a systematic 
increase in the power spectrum. For example in the $200<\ell<800$ range,
the new combined V and W band window functions are 1.5\% lower
($\approx 1.5\sigma$) than the same combination for year one, 
justifying the assumptions 
and treatment in the year-1 release. We retain nearly the same uncertainties 
on the window function that were given for year-1.

\begin{figure*}
\epsscale{0.8}
\plotone{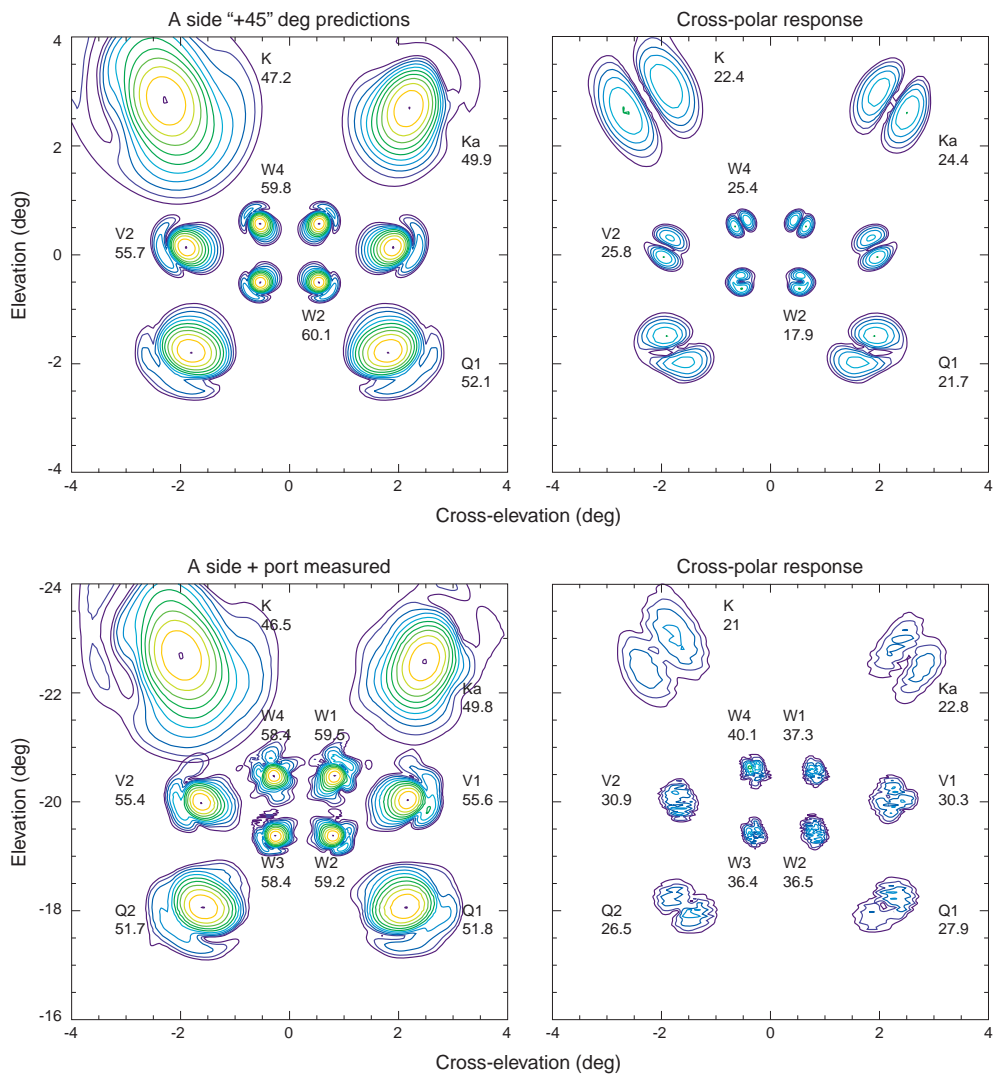}
\caption{Predicted (top) and measured (bottom) focal plane for the A side
for the co- and cross-polar beams. The contours are spaced by 3~dB
and the maximum value of the gain in dBi is given next to selected beams.
The measurement was done in the GEMAC (Goddard ElectroMagnetic 
Anechoic Chamber) beam mapping facility at NASA/GSFC. For both the predictions
and measurements, measurements at twelve frequencies across each passband
are combined using the measured radiometer response.
The difference between the predictions and measurements are
due to reflector surface deformations with an \rms of 0.02~cm.
 In flight, deformations are larger
\citep{page/etal:2003b} and require the modeling described 
in the text. In K and Ka bands, the cross-polar patterns 
are clearly evident at the predicted levels. For the other bands, the polarization 
isolation of the orthomode transducer dominates over the 
optical cross-polarization leading to a higher cross-polar gain.
This beam orientation is for an observer sitting on \wmap~ 
observing the beams as projected on the sky.}
\label{fig:focalplane} 
\end{figure*}

The co- and cross-polarization beam profiles for the A side 
are shown in Figure~\ref{fig:focalplane}. These update previous 
versions of the figures by including the cross-polar response
and data further into the tail of the beams.  The figures are 
based on pre-flight data, though the general agreement between 
the projections and measurements 
of Jupiter is excellent. The predicted and measured cross-polar 
patterns are in general agreement. 

The cross polar response of the main beam can be parameterized as a combination of a 
rotation of the linear polarization reference axis around the line of sight,
and a coupling to the Stokes V component. The relative magnitudes of these two components depends 
on the phase of the cross-polar coupling. The two outputs from each feed horn are coupled 
to different radiometers, so \wmap \ is not directly sensitive to this phase.
While neither of these components can produce an errant polarization signal from an unpolarized input signal, 
they do affect the polarization measurement. The rotation of the polarization reference axis from their design
directions was measured to be less then $1\ddeg5$ during ground testing.
The effect of the coupling to the Stokes V component is a reduction in the sensitivity to the
Stokes Q and U components, presuming no Stokes V signal is present on the sky. The largest cross polar response is
in W-band (-22~dB), corresponding to, at most, a 0.6\% reduction in polarization sensitivity relative to the 
temperature calibration. At the levels indicated, neither of these effects alter the polarization measurements
significantly. More details can be found in \citet{page/etal:prep}.

\begin{figure*}
\epsscale{1.1}
\plottwo{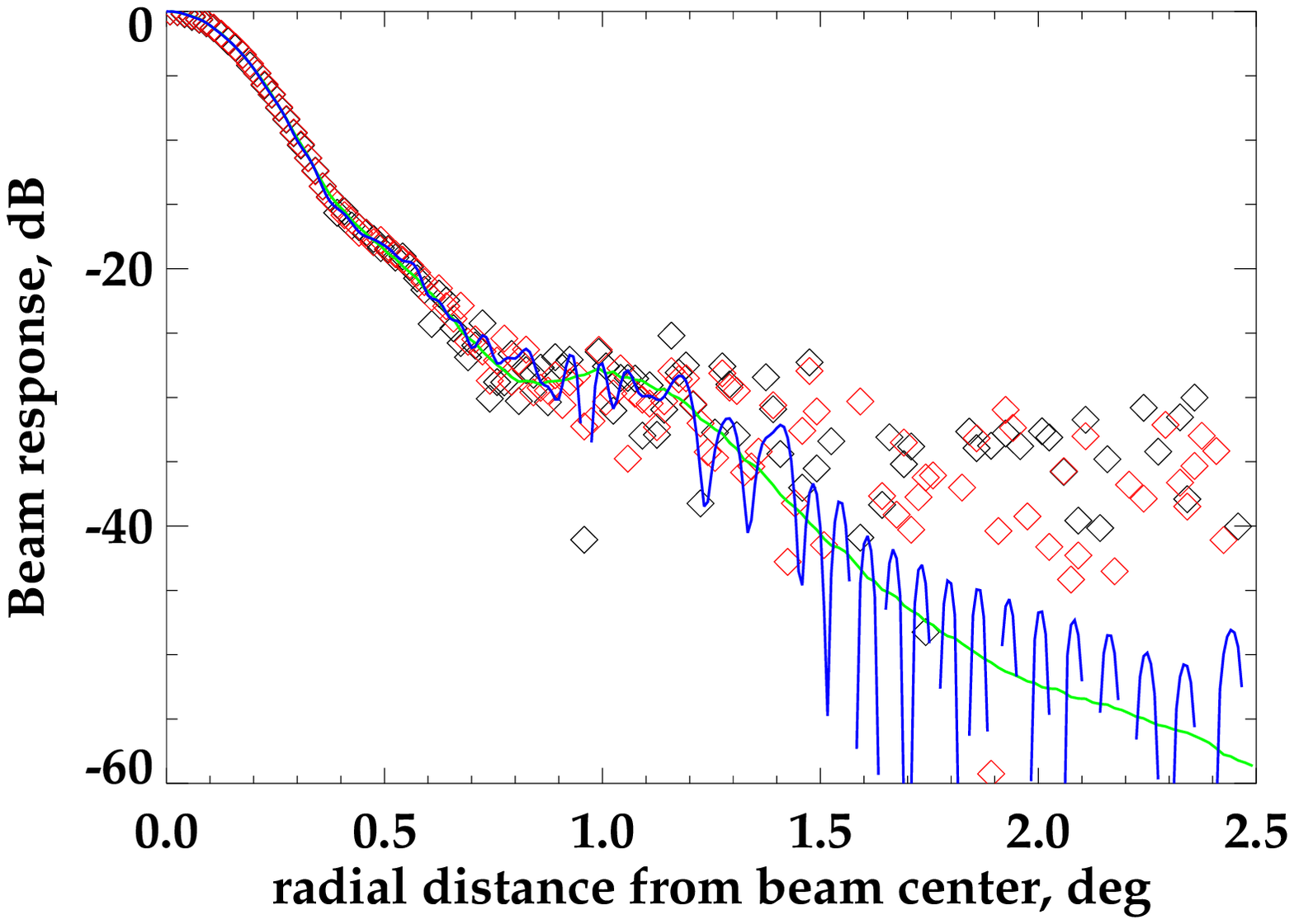}{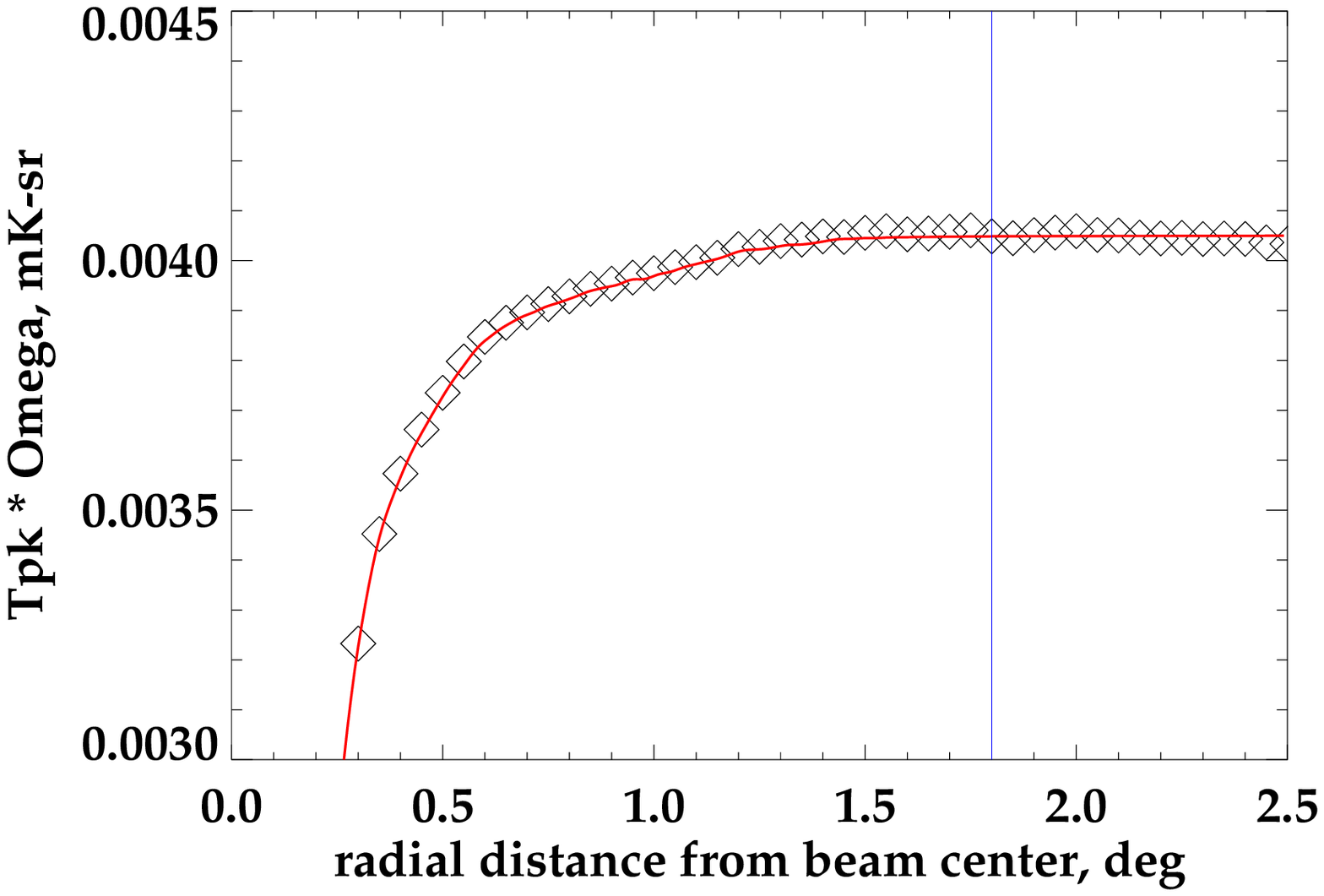}
\caption{{\it Left:} Profile of the A-side V2 band beam comparing year-1 (black),
3-year combined (red), the model of the beam (green), and the 
Hermite fit to the beam profile (blue). 
The noise level of the data is apparent at the -30 dB level.
The year-1 cutoff radius was $1\ddeg8$.
The improvements to the window functions come from understanding
the beam in the -25~dB to -30~dB region. {\it Right:}
The accumulated solid angle as a function of radius for
A-side V2. The red line in this panel corresponds to the 
model (green) in the left panel. The vertical line corresponds to the cutoff radius used
in the year-1 analysis. For this release, the integral is extended to 
$2\ddeg5$.}
\label{fig:v2profile} 
\end{figure*}

\begin{figure*}
\epsscale{1.1}
\plottwo{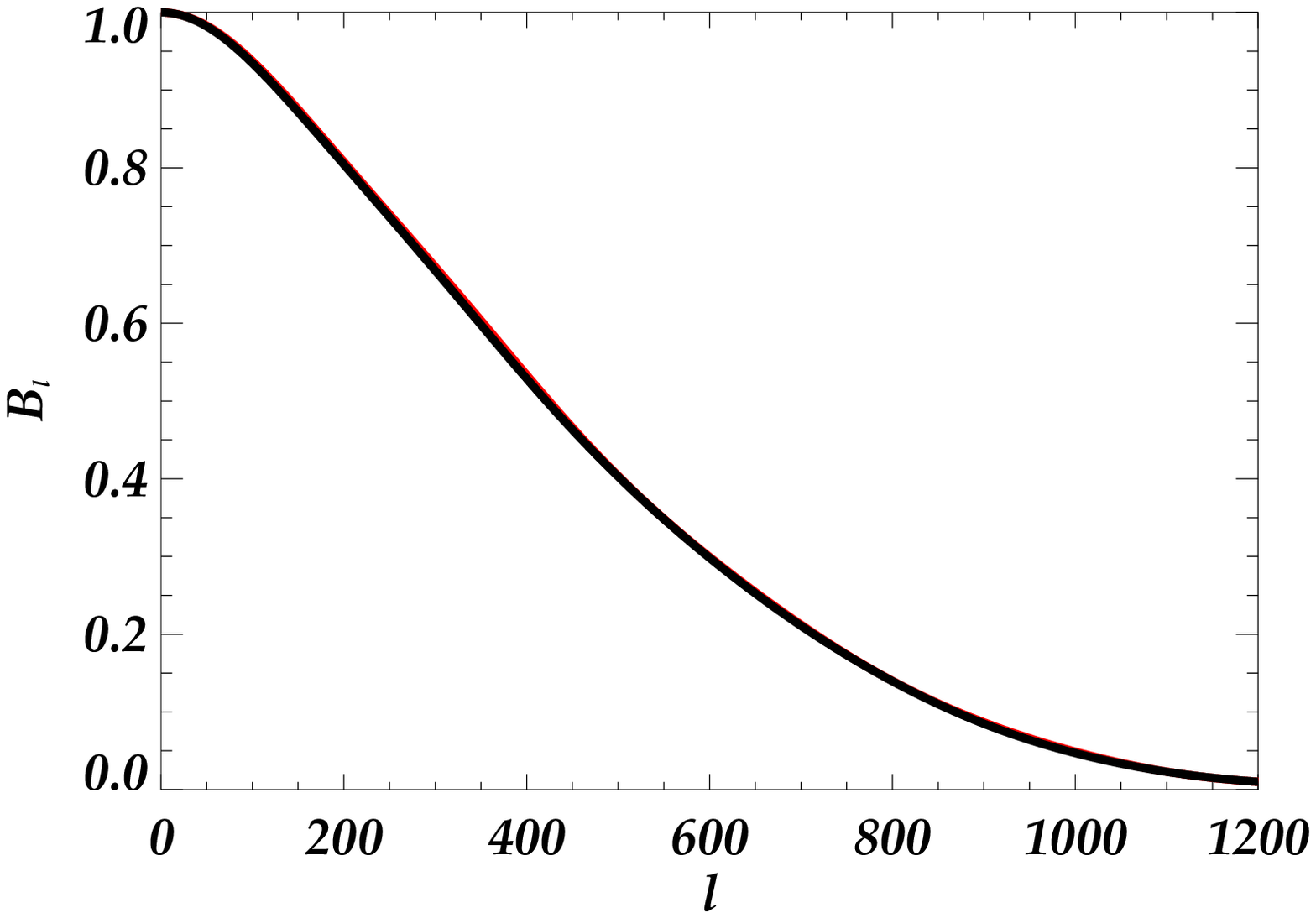}{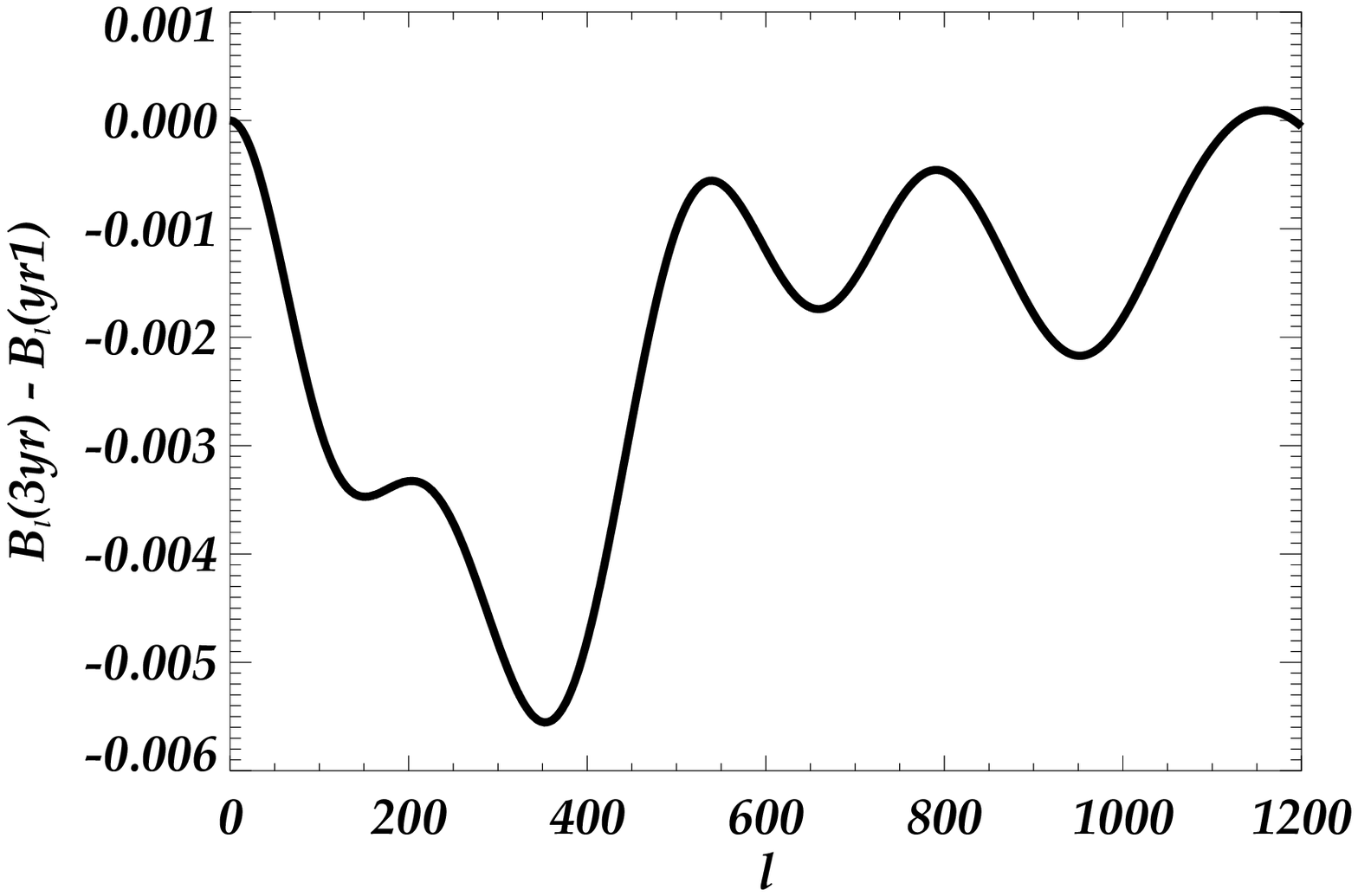}
\caption{The beam transform for the A side of the V2 DA, comparing first-year and 3-year results. The top panel shows the 
beam transform from both the first year (red) and 3-year (black) analysis --- they 
are nearly indistinguishable at this level. The bottom panel shows the difference between the beam transforms. Note that
the 3-year beam transform is lower than the first year transform near $\ell \approx 200$ by about 0.5 \%. This corresponds to 1\%
in the window function as discussed in \citet{hinshaw/etal:prep}.}
\label{fig:v2beamxform} 
\end{figure*}

The beams of the V2 DA are used to illustrate the salient aspects of 
the beam analysis.
Figure~\ref{fig:v2profile} shows the beam profile for 
year-1 and the three years combined. It also shows the accumulated solid angle as
a function of angle from the main beam axis. Figure~\ref{fig:v2beamxform} 
shows the window function derived from the profile.  
A number of features are evident. 
1) The profile is stable and is the same between years. 2) There are 
contributions to the solid angle at the -20~dBi level 
(the forward gain is 55~dBi), 
near the noise floor
of the measurement. 3) The new beam transforms are slightly different 
from the old ones.

The beams are modeled using a code based on the DADRA 
(Diffraction Analysis of a Dual Reflector Antenna)
physical optics routines \citep{YRS:DADRA}. The 
deviations from an ideal 
beam shape can be explained as deformations of the reflectors.
The shape of the primary is parametrized with a set of 122 Fourier modes
on a rectangular grid and the shape of the secondary with 30 modes. 
Based on pre-flight measurements of the cold optics,
the surface correlation length is $\approx10$~cm, which is just 
resolved by our basis functions.
A conjugate-gradient least squares method is used to solve for 
the shape of the primary and secondary reflectors. The parameters of the 
fit are the amplitudes of the modes. 
The model simultaneously solves for all ten beam profiles
and takes into account the measured passbands. For simplicity we average 
over polarizations. Though the input can be thought of 
as 10 Jupiter maps with $10^5$ net pixels, the fitting is done in the time
ordered data to 
obviate complications associated with map pixelization. The actual fit
is a multistep process. At first just a few modes on the primary are 
considered. After an approximate solution is found, more 
modes on the primary are added, and finally the modes on the 
secondary are added. The solution is annealed and the resolution 
of the model is adjusted as the fitting progresses. So far,  
we have only run the model on the A side.

The fit method works because of the wide range of frequencies, the 
high signal to noise ratio, the angular resolution of the measurement, 
and the large sampling of the focal plane. Related methods generally employ
some sort of sampling of the phase of the wave front--- 
e.g., Out of Focus (OOF) holography \citep{nikolic/richer/hills:2002}---
which of course cannot be done with \WMAP.

\begin{figure*}
\epsscale{0.4}
\plotone{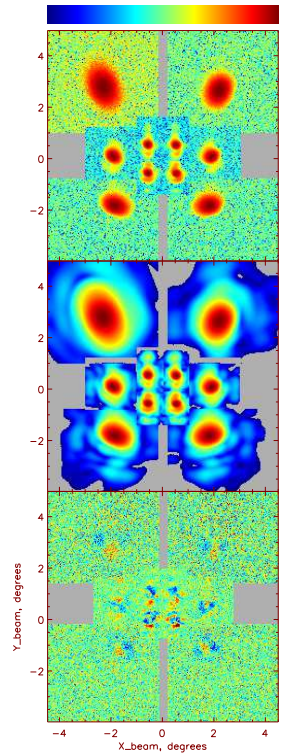}
\caption{Beams in the \wmap~focal plane. The top panel 
shows the measured beams, the middle panel shows 
the beam model, and the bottom panel shows the 
residuals. In the top two panels,
each beam is scaled to its maximum, set to 0~dB (red), and then plotted
logarithmically to a level of -40~dB (blue). For the bottom panel each beam's
residual is shown  linearly as ${\rm 100(data-model)/beam~peak}$. 
The scales are $\pm 10$\% for K; $\pm 5$\% for Ka,Q1,Q2; $\pm 3$\% 
for V1,V2; and $\pm 2.5$\% for W. 
\label{fig:modelresid} }
\end{figure*}

The residuals of the model are shown in Figure~\ref{fig:modelresid}.
These should be compared to Figure~3 of \citet{page/etal:2003} which 
shows an earlier version of the model for the A side.
For the best fit, the reduced $\chi_\nu^2=1.22$ (with roughly $10^5$ pixels). 
In other words, the model fits
to near the noise of the measurement with just 152 parameters. 
One measure of the accuracy of the model comes from the comparison 
between the modeled and the measured beam solid angle as shown 
in Table~\ref{tab:beams}. Excluding W2,
the \rms deviation is 1.6\%; the W2 solid angle is predicted to be 4.8\%
smaller than is measured. This gives us confidence that we understand 
the beams. However, as can be seen in 
the figure, there are small residual features that the model does not 
capture. These features occur mostly in the steep 
parts of the profiles corresponding to the deformations with the greatest length scales. 
Nevertheless, the model is successful at unveiling 
faint large scale components of the beam. For example, a -26~dB Q-band lobe
located $1\fdg2$ below Q2 was predicted by the model before it was 
found in the data. The model's effectiveness lies in the fact that it uses 
measurements with a high signal-to-noise ratio at all frequencies to find 
the shape of the reflector and thereby predict the beam shape 
many beamwidths off the beam axis at the noise floor of the measurements.

Figure~\ref{fig:v2profile} shows a comparison between the measured and modeled
beam. The model captures the main features of the beam though there
are some discrepancies at the -20 dB level. Most importantly, the model
reveals that the beam profile continues to drop with increasing radius. 
This is simply a result of the $\approx 10~{\rm cm}$ correlation length of the 
surface deformations. With the new model there is no longer 
a need to impose a cutoff radius as was done 
in \citet{page/etal:2003}. 

The window functions are computed from the symmetrized beam profiles following 
the Hermite method in \citet{page/etal:2003}, although 
the method is modified to take advantage of the modeling. 
The range of spacecraft orientations corresponding to the observational data contained in each sky map pixel  
depends on the ecliptic latitude of the pixel. For pixels near the ecliptic poles the observations
occur nearly uniformly for all rotations of the spacecraft around each beam line-of-sight,
effectively symmetrizing the beam. At lower ecliptic latitude the range of rotations is reduced. 
The use of the symmetrized beam profile approximation is therefore  very good near the ecliptic 
poles, but worsens at lower ecliptic latitude. The least symmetrization occurs in the ecliptic 
plane. For beams in the ecliptic plane window function errors arising from use of the symmetrized beam approximation 
are less than 1\% for $\ell < 600$ in V-band and W-band. 
In Q-band the errors are somewhat larger due to the larger beam ellipticities, with window functions 
error less than $1\%$ for $\ell < 300$. The uncertainties in the window function
determinations arising from incomplete beam symmetrization are included in the final window function uncertainties.
More details regarding beam symmetrization can be found in \citet{page/etal:2003b}.
 
Figure~\ref{fig:v2profile} shows one such fit 
for the symmetrized radial profile of the V2 A-side beam with a Hermite expansion of order 170. 
To remove some of the sensitivity of the 
window functions to the noise tail, hybrid beams are constructed.
The hybrid method retains the high signal to noise
observations of the main beam and replaces the measurement
with the model in regions of low signal.
For the A side, the hybrid beams are constructed by (a) scaling
the model to measured peak height, (b) choosing a threshold
level, $B_{\rm thresh}$ (Table~\ref{tab:beams}), and (c)
using the measurement above $B_{\rm thresh}$ and the scaled
model below it. The hybridization is done with the full two dimensional
profile in the TOD before 
the symmetrization. Since a complete model of the B side
does not yet exist, the A-side profile is translated and rotated 
to the B side focal plane, scaled to the
peak of the B-side data, and interpolated to replace B side 
observations in the TOD when they are less than $B_{\rm thresh}$.  
For both A and B sides the Hermite fits of order $n=170$ 
are made to the symmetrized
beam profiles. Analytic and numerical models  
show that the assumption of symmetrization is sufficient
except in Q band where a 4\% correction in made for $\ell>500$
\citep{hinshaw/etal:prep}.

The window functions are available with the data release. In general
they follow those in Figure~4 of \citet{page/etal:2003b}.
At low l, the windows for polarization are negligibly different
than those for temperature. At $\ell\approx 500$, the polarization windows 
can differ from the temperature windows by a few percent because 
the effective central frequencies for polarization are different 
from those for temperature. Thus, the effective beams for polarization
are different from those for temperature. We do not take this effect 
into account yet, as it is negligible at the current levels of sensitivity.

\begin{deluxetable}{cccc}
\tablecaption{Parameters of the A-side Beam Model}
\tablehead{\colhead{Band} & \colhead{$B_{\rm thresh}~$(dBi) }& \colhead{ Below peak (dB)} & \colhead{$\Omega_{\rm meas}/\Omega_{\rm mod}$}}
\startdata
K  & 17  & -30 & 0.999  \cr
Ka & 17  & -32 & 0.982  \cr
Q  & 18  & -33 & 0.971  \cr
V  & 19  & -36 & 0.999  \cr
W  & 20  & -38 & 1.018  \cr
\enddata
\tablecomments{``Below peak'' is the forward gain minus $B_{\rm thresh}$.
It is the level in dB below the peak at which the measurement
is replaced by the model. $\Omega_{\rm meas}/\Omega_{\rm mod}$ is
the measured beam solid angle ( from observations of Jupiter) divided by the 
modeled beam solid angle.}
\label{tab:beams}
\end{deluxetable}

\begin{figure*}
\epsscale{1.0}
\plotone{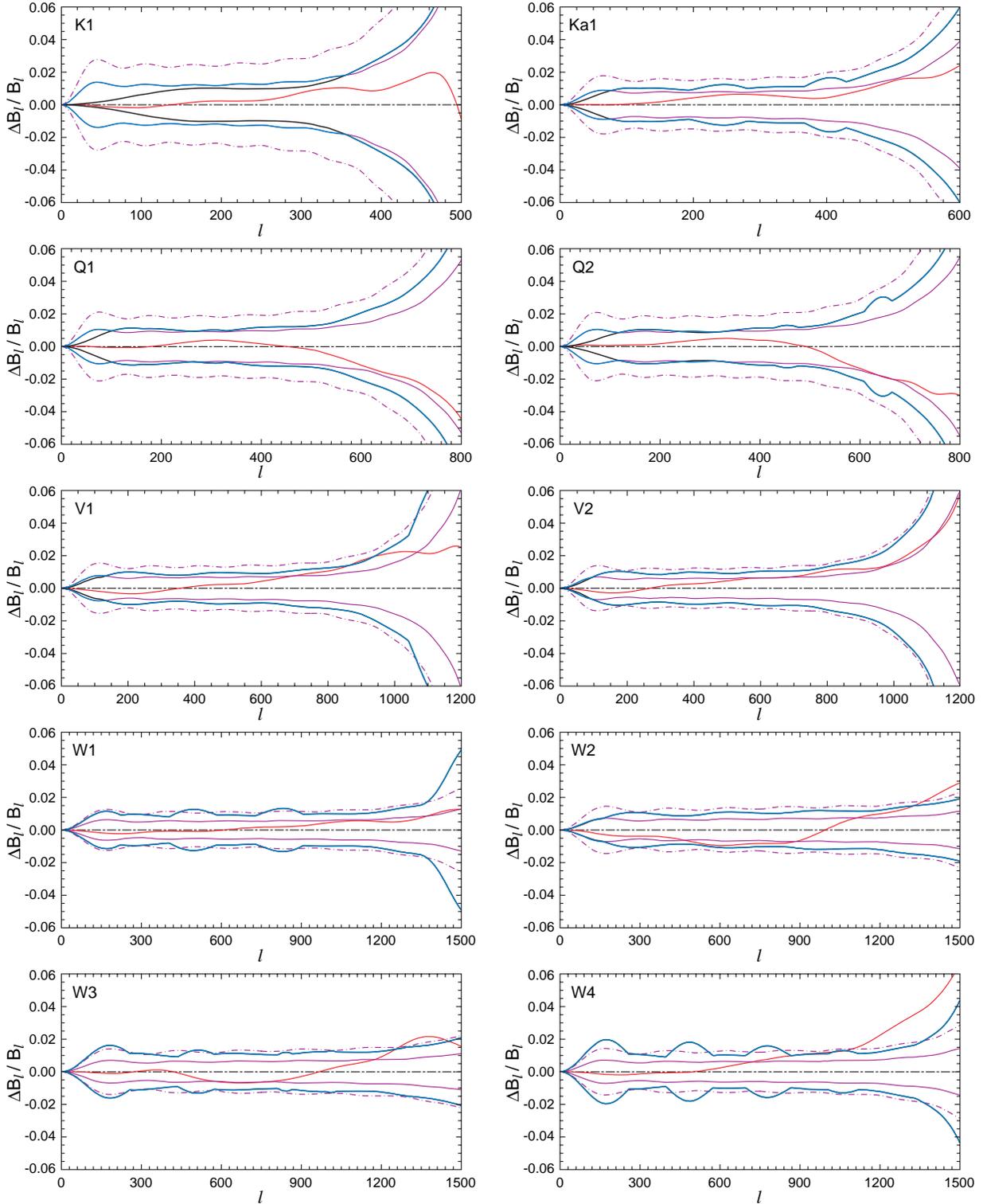}
\caption{ The uncertainties of the beam transform functions
for all bands. The uncertainties of the window functions 
are double these. The cyan lines are the new uncertainties.
The black line is the uncertainty for year-1. The inner and outer magenta lines
are the formal $1\sigma$ and $2\sigma$ uncertainties for the 
Hermite fits. The thin red line indicates the fractional 
difference between the Hermite and Legendre polynomial beam transforms.
\label{fig:winfunceff}} 
\end{figure*}

The uncertainty in the window function is 
ascertained by changing the criterion for 
merging the model and the measurements, changing the cutoff in the 
Hermite fit, comparing the Hermite and Legendre polynomial
methods, and propagating the formal errors from the fit. The
adopted uncertainties, along with a comparison to year-1, 
are shown in Figure~\ref{fig:winfunceff}. The new uncertainties are 
very close to the year-1 uncertainties except at low $l$ in K through 
Q bands where the new uncertainties are {\it larger}. A typical 
window function has an uncertainty of 2-3\%. These uncertainties add 
in quadrature in the cosmological analysis. We anticipate that 
the uncertainty can be 
reduced a factor of two after the development of the B side beam 
model and with more beam maps of Jupiter.

In addition to the work described above, two end-to-end 
consistency checks are performed. 1) We check to ensure that 
within any frequency band,
all beams yield the same value for the temperature of Jupiter.
This tells us that the beam solid angles are consistently 
computed for both A and B sides. We postpone a full reanalysis 
of the Jupiter calibration and recommend using the values
in \citet{page/etal:2003}.
2) We make separate maps of the sky for the A and
B sides and compute their power spectra. We then show 
that the ratio of the power spectra is unity to within the noise limits.
This shows that the A and B side window functions are consistent.

\subsection{ Sidelobe Corrections}
\label{sec:far_sidelobes}

All radio telescopes exhibit some degree of sensitivity to 
sources outside of the main beam.  Such 
pickup is a particular concern for CMB experiments since the measured signals  are 
small and foreground emission can be large. The total beam of a telescope 
is traditionally described as the sum of two components, the main beam
and the sidelobes. Although the  demarcation between the two components
is somewhat arbitrary, the two components sum to
the total beam response of the telescope. The \wmap~ raw time ordered data, and the maps 
reconstructed from this data,
represent the sky signal convolved with the total beam response of the telescope. 

The exponential factor in the Hermite expansion of the symmetrized beam profiles (see \S~\ref{sec:beams_and_windows}) 
and the  order of the polynomial fit determine
the maximum radius for which the Hermite expansion accurately models the beam pattern. Outside of this radius the 
Hermite expansion quickly decays. The radius at which the Hermite expansion decays, $r_{\rm H}$,  provides a natural
point at which to separate the main beam and sidelobe response: the window function derived from the Hermite expansion
represents the beam for $ r < r_{\rm H} $ and the sidelobes account for the remainder of the beams. Ideally for 
$r < r_{\rm H}$, the sidelobe map would contain residuals between the true beam pattern and that parameterized
by the Hermite expansion. However, since the main beam is modeled to levels below the noise floor of the measurements
we have no way to determine these residuals. We therefore zero the area of the sidelobe maps for $r < r_{\rm H}$.
The values of $r_{\rm H}$ are presented in Table~\ref{tab:intensity_processing1}.
These sidelobe maps, when convolved with various input signal maps, are used to implement  corrections for the
sidelobe pickup in the TOD as described in \S~\ref{sec:TOD_archive_production}. These maps differ from those 
contained in the first year data release in two aspects: 1) The region within a radius of $r_{\rm H}$ around 
each line of sight direction has been set to zero, and 2) The region of the W-band maps derived from the 
preflight sidelobe mapping has been scaled down by $\approx  5~{\rm dB}$, based on a re-evaluation of the 
calibration of those measurements. These sidelobe maps are available 
with the data release.

\begin{deluxetable}{ccc} 
\tablecaption{\wmap~Beam and Recalibration Factors \label{tab:intensity_processing1}}
\tablecomments{This Table lists recalibration applied to the time ordered 
data to compensate for calibration biases introduced by ignoring effect of the beam sidelobes
during the fitting of the calibration solution. Also presented are the maximum radii for which the
Hermite expansions of the radial beam profile are considered accurate.}
\tablecolumns{3}
\tablehead{
\colhead{DA}&
\colhead{$r_{\rm H}$}&
\colhead{Recalibration factor}\\ 

} 
\startdata
K1    & 6\ddeg1 & 1.0151 \\
Ka1   & 4\ddeg6 & 1.0047 \\
Q1    & 3\ddeg9 & 0.9971 \\
Q2    & 3\ddeg9 & 0.9975 \\
V1    & 2\ddeg5 & 1.0009 \\
V2    & 2\ddeg5 & 1.0011 \\
W1    & 1\ddeg7 & 1.0043 \\
W2    & 1\ddeg7 & 0.9985 \\
W3    & 1\ddeg7 & 0.9985 \\
W4    & 1\ddeg7 & 1.0033 \\
\enddata
\label{tab:intensity_processing2}
\end{deluxetable}

\section{DATA PROCESSING}\label{sec:data_proc}
The 3-year \wmap~ sky maps were processed as three independent data sets, each containing 1 year of
observational data. The maps were later combined to produce various data products, including a three year combined map.
The processing used to produce sky maps differs in several aspects from that used
to produce the year-1 maps. The most significant change is that the 3-year maps,
comprising Stokes I, Q and U components, are the maximum likelihood estimates of the sky map given the radiometer
``1/f'' noise characteristics. The year-1 maps were unbiased representations of the sky signal, but were sub-optimal
in terms of noise since they were produced from a pre-whitened TOD archive~\citep{hinshaw/etal:2003b}.
 Other significant processing 
changes include applying corrections for sidelobe pickup to the TOD for all DAs, inclusion of an 
estimated polarization signal in the fitting of the hourly baseline solution, and elimination of a 
weighting factor based on a radiometer noise estimate originating from the radiometer 
gain model. ( see \S~\ref{sec:gain_model}~)

\begin{figure*}
\epsscale{1.0}
\plotone{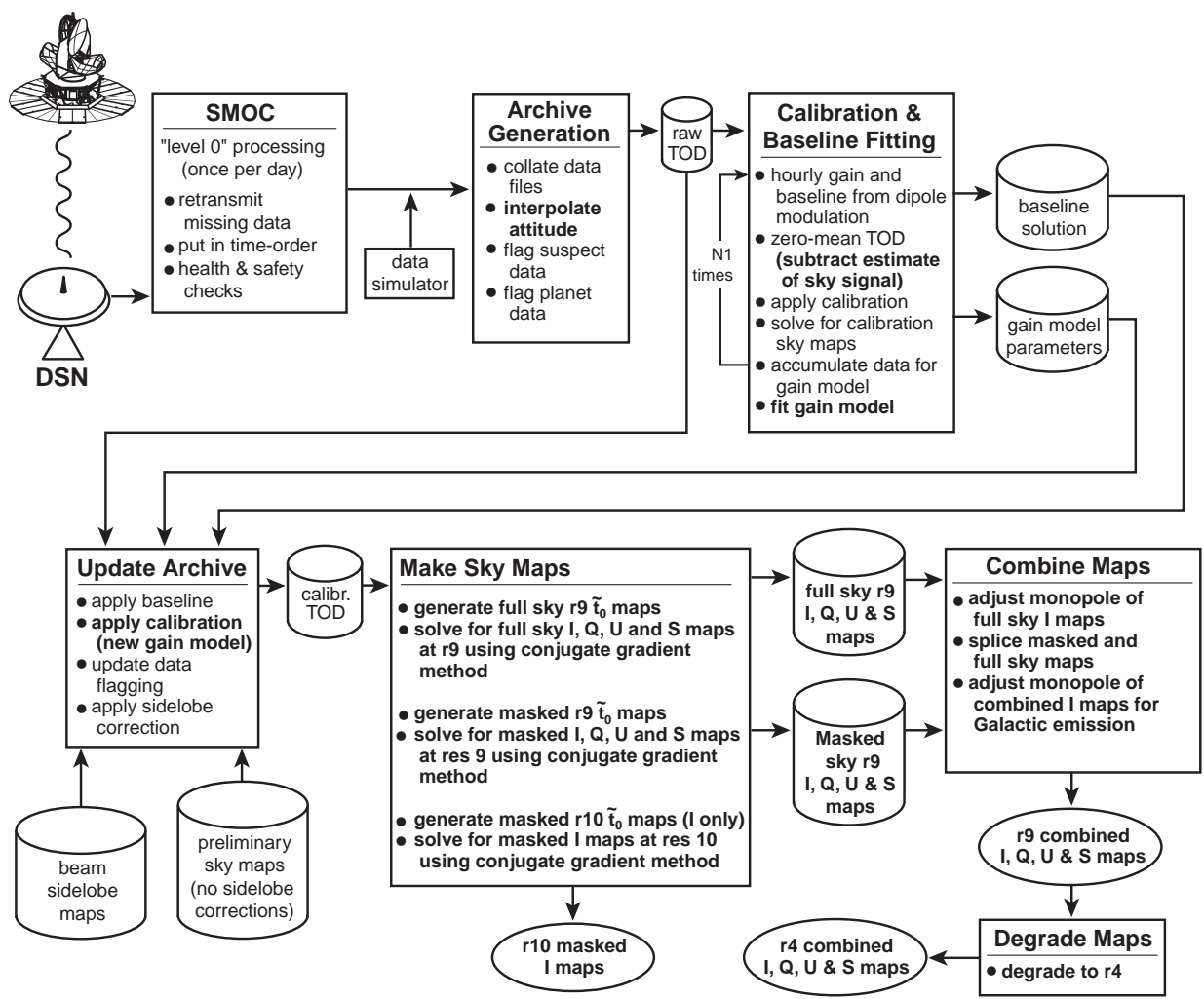}
\caption{ Schematic overview of the 3-year \wmap~sky map processing pipeline. Substantive
changes from the year-1 processing \citep[Figure 1]{hinshaw/etal:2003b} are indicated in boldface.}
\label{fig:updated_hinshaw_fig_1} 
\end{figure*}

The current data processing flow is outlined in Figure~\ref{fig:updated_hinshaw_fig_1}. Note that in the
\emph{Make Sky Maps} block three sets of maps are produced for each year of data. The first of these are full sky maps
produced at HEALPix r9. \wmap~utilizes the HEALPix\footnote{see http://healpix.jpl.nasa.gov} 
pixelization scheme and designates map resolution with the notation r4, r5, r9 and r10,  corresponding to 
HEALPix $N_{\rm side}$ parameter values of 16, 32, 512 and 1024,  and approximate pixel side dimensions of $3\ddeg7$,
 $1\ddeg8$, $0\ddeg11$ and $0\ddeg06$
respectively. 

These full sky maps suffer from small processing artifacts associated with observations
in which one of the telescope beams is in a region of strong Galactic emission 
while the other is in a low Galactic emission region. 
These artifacts arise from slight errors in the
radiometer gain determination and pixelization effects, resulting in errant signals in some
high Galactic latitudes regions used in  CMB analyses.
The year-1 map processing eliminated this problem by only updating the value in the pixel with the high Galactic emission
for such observations in the iterative map making procedure~\citep{hinshaw/etal:2003b}. Adapting this 
asymmetric masking to the conjugate gradient 
processing used to produce the current maps is not straightforward. Instead, a separate set of  r9 maps, 
termed \emph{spm} maps, is produced  using a symmetric processing mask.  These spm maps omit all 
observations for which either beam falls into a 
high Galactic emission region. The processing mask used to identify the high emission regions excludes $5.7\%$ of the pixels, and is
available with the data release.
These spm maps do not suffer the aforementioned problem, but contain no data in the high Galactic emission regions. 
Data from the full sky maps are used to fill the unobserved regions of the spm maps. 
Details of this process are presented in \S~\ref{sec:final_sky_map_production}.

The last set of maps listed in the \emph{Make Sky Maps} block of Figure~\ref{fig:updated_hinshaw_fig_1} 
is produced at  r10 and contains only spm intensity maps. These maps are used to evaluate the high-$\ell$
temperature power spectra and are produced at higher resolution to minimize pixelization effects. To speed computation these
maps were produced using only intensity information, so no corresponding high resolution polarization maps are available.

The implementation of the entire data processing pipeline has been exhaustively verified through
numerous simulations of TOD with the non-idealities described in \S~\ref{sec:radiometer_non-idealities}, both with and without inclusion of instrument noise. In the case 
excluding instrument noise the pipeline has been shown to reproduce the simulated input maps to sub-nanoKelvin
accuracy. Simulations with instrument noise have been shown to produce unbiased estimates of
the input simulated sky maps. The subsequent discussion follows the notation of \citet{hinshaw/etal:2003b}
and updates the description of the data processing therein.

\subsection{SMOC processing  and Archive Generation}
Processing of the data in the Science and Mission Operations Center (SMOC) and Archive Generation 
steps~(see Figure~\ref{fig:updated_hinshaw_fig_1}) is virtually unchanged from that used in  year-1.
The only substantive change is that corrections for the thermally induced star tracker position errors ( \S~\ref{sec:pointing})
are applied to the star tracker data in calculation of the pointing quaternions included in the raw TOD archive. 

\subsection{ Calibration and Baseline Fitting}
The calibration and baseline fitting procedures, used to calibrate the 
radiometric data based on the CMB dipole signal,
are similar to those used in the year-1 processing~\citep{hinshaw/etal:2003b}.
This is an iterative process in which both
calibration data and approximate sky map solutions are generated. Hourly estimates of the radiometer gain are made by fitting each 
1 hour segment of the TOD, corrected for estimated sky signals (excluding the CMB dipole), to the sum of a template 
dipole signal and a baseline term that accounts for very slow drifts in the radiometer outputs. The template dipole signal is derived 
from \wmap~ velocity and attitude information and the measured barycenter dipole from the first-year \WMAP~results. Improved sky maps 
are then generated using the updated baseline and hourly gain solutions, which in turn are 
used to produce improved baseline and gain solutions. The 3-year processing is  an improvement over the 
1-year processing in that it corrects the TOD for estimated sky signals based on the Stokes I, Q and U 
components before fitting for the gain and baseline, whereas in the year-1 processing only the Stokes 
I signal was removed from the TOD before the baseline and gain solution were fit. 

The \wmap \ scan strategy ensures that the Stokes I sky signal components average
very nearly to zero over a 1 hour precession period. However, the orientation of the polarization axis 
of the feed horns and scan pattern can transform certain 
sky \emph{polarization} signals into TOD signals with very long periods, extending to many hours or days.
 The baseline fitting routine forces the
mean of the signal corrected TOD to zero on time scales of one hour and longer. Removing the polarization signal before
performing the baseline fitting ensures that the polarization signals contained in the TOD remain unbiased.

After the hourly gain and baseline solutions are converged they are used to fit the parameters of the 
improved gain model described in \S~\ref{sec:gain_model}.

\subsection{Calibrated TOD Archive Production} \label{sec:TOD_archive_production}
The production of a final calibrated TOD archive involves two major steps: first the gain and baseline 
are applied to the uncalibrated TOD using the parameters determined in the calibration and baseline fitting procedure. 
Two corrections related to sidelobe response are then applied to this initial calibrated archive. 

The first sidelobe correction
simply removes an estimate of the signal arising from sidelobe pickup. This signal is calculated by convolving the
Stokes I sidelobe response maps (\S~\ref{sec:far_sidelobes}) for each DA with 
preliminary Stokes I sky maps using the 
technique of ~\citet{wandelt/gorski:2001}. The input sky maps contain three components: 1) An estimated CMB 
+ Galactic signal obtained from a preliminary set of maps
produced without sidelobe corrections, 2) a fixed barycenter CMB dipole signal, and 
3) an annually modulated CMB dipole 
signal from Earth's motion relative to the solar system barycenter. 
These corrections apply to the Stokes I signal only and are therefore applied equally 
to the data from both radiometers comprising each DA. This correction is applied to the TOD on a sample-by-sample basis.
 The effects of polarized sidelobe pickup are small 
~\citep{barnes/etal:2003} and are not corrected in this data release.

The second sidelobe related correction compensates for a small calibration error introduced in 
the initial calibration process by multiplying the sidelobe corrected TOD archive by an overall scale factor.
 The uncalibrated TOD archive used to fit the gain and baseline 
solution contains signal arising from both the main  beam and the sidelobe response of the
telescope optics to the true sky signal. However, the template dipole signal used to 
fit the gain is based on ideal pencil beams in the boresight directions of each telescope beam. This
approximation can lead to biased gain measurements since it ignores the component of the TOD arising from sidelobe pickup
of the sky signal. This effect has been
studied through simulations in which simulated TOD, based on the sidelobe response maps, is input to the hourly gain fitting
algorithm to determine the level of bias originating from the sidelobe signals. The bias is found to be relatively small and
varies only slightly over the course of a year as the  precession axis of the \wmap~scan pattern sweeps across the sky.
Table~\ref{tab:intensity_processing1} lists the values of these recalibration factors.  
A detailed description of this correction is presented in Appendix~\ref{app:sidelobe_corr}.
 
\subsection{Maximum Likelihood Map Solution}
\label{sec:cgmaps}
As described in \citet{hinshaw/etal:2003b}, the \wmap~ time ordered data may be written as
\begin{equation}
{\bf d} = {\bf M t} + {\bf n},
\end{equation}
where $\mathit{\bf M}$ is the mapping matrix that transforms a sky map, $\mathit{\bf t}$, into
discrete samples,  $\mathit{\bf d}$, and $\mathit{\bf n}$ is the noise added to each sample by the radiometer. The map making problem
consists of generating an estimated sky map given $\mathit{\bf d}$ and the statistical properties of the 
noise. Taking the noise description as 
\begin{eqnarray}
\langle{\bf n}\rangle &=& 0, \\ 
\langle{\bf nn^T}\rangle &=& {\bf N},
\end{eqnarray}  
it is well known that the maximum likelihood estimate of the sky map, $\mathit{\bf \tilde{t}}$,  is
\begin{equation}
{\bf \tilde{t} } = ({\bf M}^T {\bf N}^{-1} {\bf M })^{-1} \cdot ({\bf M}^T {\bf N}^{-1} {\bf d}). \label{eqn:map_soln}
\end{equation}
In the case of a single differential radiometer and sky maps comprising only Stokes I,
 ${\bf d}$ is a data vector comprised of $\mathit{N_t}$ time ordered data elements,
${\bf \tilde{t}}$ is a sky map of ${N_p}$ pixels, ${\bf M}$ is an ${N_t \times N_p}$ matrix
and ${\bf N^{-1}}$ is an ${N_t \times N_t}$ element matrix. The matrix ${\bf M}$ is sparse, each row corresponding to one
observation, and each column corresponding to a map pixel. For an ideal differential radiometer
each row contains a value of $+1$ in the column corresponding to the pointing of the A-side beam for a given observation, and
a value of $-1$ in the column corresponding to the B-side beam. 
Given these matrices it is possible to produce maps from the \wmap~data set
using  equation~(\ref{eqn:map_soln}).
The second term on the right hand side
 of equation~(\ref{eqn:map_soln}) may be evaluated directly, while multiplication by the first term, $({\bf M^T N^{-1} M })^{-1}$, 
may be performed using a conjugate gradient iterative technique.
It is straightforward to generalize this technique to process polarization maps. 

As described in \citet{bennett/etal:2003}, the signal from each telescope beam (A-side and B-side) is separated into two orthogonal linear
polarizations by orthomode transducers at the base of the feed horns. Each pair of feed horns is associated with two
differential radiometers comprising the DA. One linear polarization from each feed horn
is fed into one differential radiometer, while the orthogonal set of linear polarizations are fed 
to the second radiometer comprising the DA. Details of the alignment of the polarization axis and beam boresights
were presented in \citet{page/etal:2003}.

The outputs of radiometers ``1'' and ``2'', ${\bf d}_1$ and ${\bf d}_2$, for each observation 
correspond to sky signals
\begin{equation}
{\bf d_1 } = {\bf i}(p_{\rm A}) + {\bf q}(p_{\rm A}) \cos 2 \gamma_{\rm A} +   {\bf u}(p_{\rm A}) \sin 2 \gamma_{\rm A} -
  {\bf i}(p_{\rm B}) - {\bf q}(p_{\rm B}) \cos 2 \gamma_{\rm B} -   {\bf u}(p_{\rm B}) \sin 2 \gamma_{\rm B}
\end{equation}
and
\begin{equation}
{\bf d_2 } = {\bf i}(p_{\rm A}) - {\bf q}(p_{\rm A}) \cos 2 \gamma_{\rm A} -   {\bf u}(p_{\rm A}) \sin 2 \gamma_{\rm A} -
  {\bf i}(p_{\rm B}) + {\bf q}(p_{\rm B}) \cos 2 \gamma_{\rm B} +   {\bf u}(p_{\rm B}) \sin 2 \gamma_{\rm B}.
\end{equation}
Here ${\bf i}(p_{\rm A})$, ${\bf q}(p_{\rm A}) $ and ${\bf u}(p_{\rm A})$ are the 
Stokes  I, Q  and U sky signals at sky position $p_{\rm A}$. The 
polarization angle of radiometer ``1'' with respect to the sky reference direction for sky position $p_{\rm A}$
is $\gamma_{\rm A}$ as described in \citet{hinshaw/etal:2003b}. Variables with subscript B are the corresponding
values for the sky position observed by the B-side beam. 

The mapping matrix ${\bf M}$ is generalized to include polarization processing by expanding it to $ 2 N_{\rm t} \times 3 N_{\rm p}$ matrix
where the number of rows has been doubled since there are two data values from each observation, and the number of columns has
been increased to accommodate 3 maps, corresponding to the three Stokes parameters. Each row of the polarization mapping matrix has
6 non-zero elements, with $\pm 1$  in columns corresponding to the observed pixels (A-side and B-side for each radiometer ) in 
the ${\bf i}$ map and
$\pm \cos  2 \gamma_{\rm A},  \pm \sin 2 \gamma_{\rm A}$, $ \pm  \cos 2 \gamma_{\rm B }$ and $\pm \sin 2 \gamma_{\rm B}$,
in appropriate locations in the columns corresponding to the ${\bf q}$ and  ${ \bf u}$ maps. Each observation is
associated with 12 non-zero values of the mapping matrix that are distributed in two rows. The noise matrix
is similarly expanded to account for the signals from both radiometers. The noise of the two radiometers is uncorrelated, so it 
may be described by the relations,
\begin{eqnarray}
\langle{\bf n}_1\rangle = \langle{\bf n}_2\rangle = 0, \\ \label{eq:noise_cross_corr}
\langle{\bf n}_1 {\bf n}_2 \rangle = 0, \\
\langle{\bf n}_1 {\bf n}_1^T\rangle = {\bf N}_1, \\
\langle{\bf n}_2 {\bf n}_2^T\rangle = {\bf N}_2. 
\end{eqnarray}
The full noise covariance matrix is simply a block diagonal combination of ${\bf N}_1$ and ${\bf N}_2$. 

The noise from different radiometers was verified as uncorrelated (eq.[\ref{eq:noise_cross_corr}]) by
evaluation of the cross correlation coefficient
\begin{equation}
C_{\rm 1 2} = \frac{\langle{\bf n}_1 {\bf n}_2 \rangle}{\sqrt{\langle{\bf n}_1 {\bf n}_1 \rangle\langle{\bf n}_1 {\bf n}_2 \rangle}}
\end{equation}
where $n_1$ and $n_2$ were obtained from the TOD by subtracting an estimated signal based on a combined
3 year final sky map. For all DAs, except K1, the measured value of the magnitude of $C_{\rm 1 2}$ (at zero time lag) 
was $-0.001 < C_{\rm{1 2}} < 0$. Similar results were obtained from simulations of sky signal plus noise 
which were constructed to have zero noise cross correlations. 
The small values of  anti-correlation observed arise from noise in the common sky map used to remove the sky signal 
from the  TOD of the two radiometers. For K-band, incomplete removal of the 
intense Galactic signal due to small gain errors ($\approx 0.1\%$) lead to a measured
 $C_{\rm 1 2}$ value of $\approx 0.007$, similar to values obtained from the simulation which also included gain uncertainties.

\subsubsection{ Radiometer Non-Idealities} \label{sec:radiometer_non-idealities}
If the \wmap~radiometers were ideal, solutions based on the mapping matrix described in the previous section
would reproduce the maximum likelihood estimate of the sky signal. There are two known 
instrumental non-idealities that affect the maps and are treated in the analysis:
bandpass mismatch and input transmission imbalance.

\subsubsubsection{Bandpass Mismatch} \label{sec:bandpass_mismatch}
The construction of the mapping matrix ${ \bf M}$ presented in \S \ref{sec:cgmaps}~implicitly
assumed that the microwave frequency response of the two radiometers comprising each DA were identical.
The \wmap~radiometers have slightly differing frequency responses \citep{jarosik/etal:2003b} which can cause signals with spectra
different from that of the calibration source (the CMB dipole) to be aliased into polarization maps \citep{barnes/etal:2003}. This
\emph{spurious} signal has the characteristic that it appears as a difference between the response of the two
radiometers comprising a DA, ${\bf d}_1 - {\bf d}_2$, but is not modulated with the polarization angles $\gamma_{\rm A}$
 and $\gamma_{\rm B}$ as a true polarization signal would be, since it only depends on the intensity 
and spectral index of the source region.
 This effect can be treated in the map making procedure by considering the signal from the radiometers
to originate from 4 source maps, the original 3 Stokes parameters plus a spurious map, ${\bf s}$. Including this extra term, the TOD is 
then described by the relations 
\begin{equation}
{\bf d_1 } = {\bf i}(p_{\rm A}) + {\bf q}(p_{\rm A}) \cos 2 \gamma_{\rm A} + {\bf u}(p_{\rm A}) \sin 2 \gamma_{\rm A}  + {\bf s}(p_{\rm A}) -
  {\bf i}(p_{\rm B}) - {\bf q}(p_{\rm B}) \cos 2 \gamma_{\rm B} -   {\bf u}(p_{\rm B}) \sin 2 \gamma_{\rm B} - {\bf s}(p_{\rm B})
\end{equation}
and
\begin{equation}
{\bf d_2 } = {\bf i}(p_{\rm A}) - {\bf q}(p_{\rm A}) \cos 2 \gamma_{\rm A} - {\bf u}(p_{\rm A}) \sin 2 \gamma_{\rm A}  - {\bf s}(p_{\rm A}) -
  {\bf i}(p_{\rm B}) + {\bf q}(p_{\rm B}) \cos 2 \gamma_{\rm B} +   {\bf u}(p_{\rm B}) \sin 2 \gamma_{\rm B} + {\bf s}(p_{\rm B}).
\end{equation}
Note that for the combination of radiometers outputs corresponding to Stokes I, ${\bf d_1} + {\bf d_2}$,  the spurious term cancels, and
for the combination corresponding to polarization signals,  ${\bf d_1} - {\bf d_2}$, it appears as a scalar map, independent
of polarization angles $ \gamma_{\rm A}$ and  $ \gamma_{\rm B}$. Equation~(\ref{eqn:map_soln}) is easily generalized to
handle these extra terms by expanding the mapping matrix to  $2 N_{\rm t} \times 4 N_{\rm p}$, where the extra columns now correspond to the
spurious signal map. Each row of this matrix now has 8 non-zero entries, 2 corresponding to each type of map, ${\bf i, q, u}$ and ${\bf s}$.

\subsubsubsection{ Input Transmission Imbalance } \label{sec:InputTransmissionImbalance}
An ideal differential radiometer only responds to differential input signals and completely rejects common mode signals.
 Due to several effects, the \wmap~ radiometers exhibit a small response to common mode signals~\citep{jarosik/etal:2003b}. 
This response is characterized by a transmission imbalance factor, $x_{\rm im}$, with the response of the radiometer, ${\bf d}$,
to input signals $T_{\rm A}$ and $T_{\rm B}$ given by equation~\ref{eqn:loss_imbalance_def}.
The  values of $x_{\rm im}$ for the 3-year data are given in Table~\ref{tab:transmission_imbalance}. The effect can easily be
incorporated into the maximum likelihood maps solution (see eq.[\ref{eqn:map_soln}]) by modification of the mapping matrix, $\mathit{\bf M}$.
The TOD, including both the spurious map term and the transmission imbalance terms is given by
\begin{eqnarray}
{\bf d_1 } = (1+x_{\rm im})({\bf i}(p_{\rm A}) + {\bf q}(p_{\rm A}) \cos 2 \gamma_{\rm A} + {\bf u}(p_{\rm A}) \sin 2 \gamma_{\rm A}  + {\bf s}(p_{\rm A})) + \\
  (1-x_{\rm im})(-{\bf i}(p_{\rm B}) - {\bf q}(p_{\rm B}) \cos 2 \gamma_{\rm B} -   {\bf u}(p_{\rm B}) \sin 2 \gamma_{\rm B} - {\bf s}(p_{\rm B}))
\end{eqnarray}
and
\begin{eqnarray}
{\bf d_2 } = (1+x_{\rm im})({\bf i}(p_{\rm A}) - {\bf q}(p_{\rm A}) \cos 2 \gamma_{\rm A} - {\bf u}(p_{\rm A}) \sin 2 \gamma_{\rm A}  - {\bf s}(p_{\rm A}))+ \\
 (1-x_{\rm im}) (-{\bf i}(p_{\rm B}) + {\bf q}(p_{\rm B}) \cos 2 \gamma_{\rm B} +   {\bf u}(p_{\rm B}) \sin 2 \gamma_{\rm B} + {\bf s}(p_{\rm B})).
\end{eqnarray}

The corresponding modification to ${\bf M}$ is that each term involving A-side data is multiplied by a
factor  $(1+x_{\rm im})$ and those involving B-side data  are multiplied by a
factor  $(1-x_{\rm im})$.

\subsubsection{ Map Evaluation} \label{sec:map_evaluation}

The maximum likelihood map solution (eq.[\ref{eqn:map_soln}]) is evaluated in two steps. First the 
product ${\bf{\tilde{ t}}}_0 = {\bf M}^T {\bf N}^{-1}{\bf d}$, the ``iteration 0'' maps, are formed. The ${\bf \tilde{t}}_0$ 
maps are then multiplied by the $({\bf MN}^{-1}{\bf M}^T)^{-1}$ term using an iterative technique.

\subsubsubsection{ Generation  of the  ${\bf{\tilde{ t}}}_0$ Maps}
The calibrated TOD, ${\bf d}$,
represents the entire sky signal, including the CMB dipole components. To minimize numerical errors a TOD signal corresponding
to a nominal CMB dipole is removed from the calibrated TOD before evaluation of the ${\bf \tilde{t}}_0$ maps. The dipole signal 
removed from the TOD
for each radiometer includes the effects of loss imbalance based on the values given in Table~\ref{tab:transmission_imbalance}.
A CMB dipole amplitude of 3.3463 \mkelvin (thermodynamic) in the direction $(l, b) = (263\ddeg87,~48\ddeg2)$ is used to calculate
the barycentric component. A CMB monopole temperature of 2.725 K (thermodynamic)~\citep{mather/etal:1999} is 
used to calculate the annually varying component
due to the spacecraft's motion about the barycenter. The kinematic quadrupole in not removed from the TOD in the generation of the 
${\bf{\tilde{ t}}}_0$ maps.

The dipole-subtracted TOD from each radiometer must be multiplied by a filter consisting of the inverse of 
each radiometer's noise correlation 
matrix. This is accomplished by forming the inverse noise correlation function, 
\begin{equation}
N_{\rm tt}^{-1}(\Delta t) \equiv \left\{ \begin{array}{ll}
C \ \left[ \int e^{i \omega \Delta t}\left(\int e^{i \omega t'} N(t') dt'\right)^{-1} d\omega + K \right], & |\Delta t| < \Delta t_{\rm max} \\
0, & |\Delta t| \geq \Delta t_{\rm max}, 
\end{array}
\right. 
\end{equation}
where $N(t')$ is the parametrized noise correlation function as described in \S ~\ref{sec:TOD_power_spectra}.
Since $N(t')$ is constructed to be zero at lags greater than $\Delta t_{\rm max}$, $N_{\rm tt}^{-1}$ also falls to zero on the 
same time scale, so values of the filter, $N_{\rm tt}^{-1}$,  are also set to zero for lags at which $N(t')$ was set to zero. 
An example of the steps involved in forming this filter function for the W11 radiometer is shown in Figure \ref{fig:ACF_filter}.
Next, a constant $K$ is
added to the values of $N_{\rm tt}^{-1}$ for lags less than $\Delta t_{\rm max}$ to force the mean of this portion of the filter to zero.
 This ensures that very low 
frequencies, those with periods longer than the filter extent, are filtered out. Finally, the filter is normalized by an overall
multiplicative constant, $C$,  such that the value at $\Delta t = 0$ is unity. Note that none of these adjustments to $N_{\rm tt}^{-1}$ 
biases the resultant
sky maps since the same form of $N_{\rm tt}^{-1}$ is used in both terms in equation~(\ref{eqn:map_soln}). 
Tabulated versions of these filters are contained in the data release.

The radiometer noise properties are treated as stationary for each year of data, making the inverse noise matrix, 
${\bf N}^{-1}$,  circulant. The matrix multiplication, ${\bf N}^{-1}~\cdot~ {\bf d}$, may therefore be 
implemented through use of the convolution \mbox{$N_{\rm tt}^{-1} \otimes {\bf d}$}, 
and is implemented using standard Fourier techniques. Missing samples in the TOD, due either to gaps in the 
data or masking when producing spm maps, are zero-padded to properly preserve the time relationship 
between data on either side of the gaps.

Given the filtered TOD, the ${\bf\tilde{t}}_0$ maps are evaluated by accumulating the product of the corresponding 
elements of ${\bf M}^T$ and the TOD sample by sample. Cut sky maps are formed by simply
replacing the rows of ${\bf M}$ with zeros for observations where the data are masked. These ${\bf \tilde{t}}_0$ maps correspond to
the sky maps weighted by the inverse of the pixel-pixel noise correlation matrix, ${\bf \Sigma}^{-1}$,
\begin{equation}
{\bf {\tilde{t}}}_0 =  {\bf M}^T {\bf N}^{-1} {\bf d} = {\bf M}^T {\bf N}^{-1} {\bf M}{\bf t} = {\bf \Sigma}^{-1} {\bf t}   
\end{equation}
where ${\bf t}$ represents sky maps of the three Stokes parameters, I, Q and U, and a map corresponding to the 
spurious signal described in \S~\ref{sec:bandpass_mismatch}. These maps are available with the data release. Solving for the
sky maps simply requires multiplication by the pixel-pixel noise covariance matrix, ${\bf \Sigma}$.

\subsubsubsection{ Final Sky Map Production} \label{sec:final_sky_map_production}
The final step in the sky map production, multiplication of the ${\bf \tilde{t}}_0$ maps by the pixel-pixel 
noise correlation matrix, ${\bf \Sigma}$, 
is effected through  a conjugate gradient iterative technique. This method allows 
solution of the symmetric positive definite system
\begin{equation}
 {\bf Ax} = {\bf b} \label{eqn:CG_template}  
\end{equation}
for vector ${\bf x}$ given vector ${\bf b}$ and the ability to multiply an arbitrary vector by the matrix ${\bf A}$.
In this application vector ${\bf b}$ is a ${\bf \tilde{t}}_0$ map and matrix ${\bf A}$ corresponds to the inverse of the corresponding
pixel-pixel noise correlation matrix, ${\bf A} = {\bf \Sigma}^{-1} = {\bf M}^T {\bf N}^{-1} {\bf M}$. Multiplication of this
matrix by a vector representing a set of sky maps is a straightforward operation. A simulated time stream is generated
from the input maps by stepping through the archive evaluating the 16 non-zero elements  of the mapping matrix (contained in two rows) corresponding 
to each observation (the ${\bf M}$ operation). The resultant time stream is then filtered 
and re-accumulated into a new map (the ${\bf M}^T {\bf N}^{-1}$) operation using the same method previously described to 
produce the ${\bf \tilde{t}}_0$ maps. The conjugate gradient method then iteratively improves estimates of $x$, with the level of convergence being
characterized by the quantity $\epsilon$ given by the relation
\begin{equation}
\epsilon = \frac{ || {\bf b} - {\bf Ax} ||}{ || {\bf b} || } =  \frac{ || {\bf{\tilde{t}}}_0 - {\bf \Sigma}^{-1}{\bf t }||}{ || {\bf{ \tilde{t}}}_0 || },
\end{equation}
where $||x||$ is the vector-norm operator.
The iterative processing was stopped when the condition $ \epsilon \leq 10^{-8} $ was satisfied. 
\subsubsubsection{Preconditioning}
The rate of convergence of the conjugate gradient method is quite sensitive to the properties of the matrix ${\bf A}$,
with nearly diagonal forms preferable. If another tractable matrix multiplication operation can be found which approximates
multiplication by ${\bf A}^{-1}$, the rate of convergence can be greatly increased. Consider the case where the matrix
$\tilde{\bf A}^{-1}$ meets these criteria, multiplication of equation~(\ref{eqn:CG_template}) by $\tilde{\bf A}^{-1}$
yields
\begin{equation}
(\tilde{\bf A}^{-1} {\bf A}) {\bf x} = \tilde{\bf A}^{-1}{\bf b},
\end{equation}
which has the same solution as equation~(\ref{eqn:CG_template}), but the matrix product $\tilde{\bf A}^{-1} {\bf A}$
is nearly diagonal, speeding convergence of the solution. This technique was used in production of the final sky maps.
The preconditioner was implemented by separating the input map into high resolution (r9) and low resolution (r4)
components. The low resolution map was formed by re-binning the high resolution map with uniform weights. This low 
resolution map was then subtracted from the high resolution map, leaving only small angular scale information in the high resolution map. 
 The low resolution component was multiplied by the inverse of pixel-pixel noise correlation 
matrix (\S ~\ref{sec:inverse_pp_noise_matrix})
and the high resolution component multiplied by  the reciprocal of the diagonal components of the full resolution noise correlation matrix.
The two components were then recombined to produce the preconditioned map. The map production algorithms were 
carefully checked using numerous simulations
to verify that they converged to the correct solutions. It was also verified that this solution was independent of the exact form of the
preconditioner employed, although as expected the convergence rates did vary. Using the preconditioner as described, the  r9
sky map sets, comprising Stokes I, Q and U components, and the spurious map S, converged in $\approx 50-100$ iterations  to the level described above.

\subsubsection{Combining Full Sky and Cut Sky Maps}\label{sec:combined_maps}
Both full sky and cut sky versions of maps were produced for all 
three years of observations. The final sky maps for each year were produced by
using data from the full sky maps to fill in the regions of missing data in the masked maps. 
Before combining the data from these maps the
mean temperature of the full sky Stokes I and the S map required adjustment. For an ideal differential radiometer the mean
value of the I and S maps would be undetermined, corresponding to singular modes in ${\bf \Sigma}$. However, inclusion of the 
non-zero valued transmission imbalance factors, $x_{\rm im}$,  converts these previously singular modes into modes with
very small eigenvalues, indicating that they are very poorly constrained by the measurement. While no physical significance is attached
to the recovered values, allowing the mean to vary is required to achieve the level of convergence previously described
with regard to the conjugate gradient solution. The fact that the map means are poorly constrained 
indicates that estimates from the two different map processings may differ significantly due to statistical fluctuations. 

Failure to adjust the relative values
of the means would introduce an obvious discontinuity at the border of the masked regions when the maps were combined. 
The procedure used to combine the maps is as follows.
First the set of pixels which have the same number of observations in both versions of the sky maps were identified. A constant was then added 
to the full sky map such that the mean values of the previously identified sets of pixels in the full sky  map equaled the mean value of 
the same set of pixels in each corresponding cut sky map. The pixels containing no observations in the cut 
sky map were then set to the values of the corresponding pixels in the adjusted full sky maps for the I and S maps. The mean values of the
Q and U maps are well determined and were not adjusted. This procedure preserves the mean value for all four map components 
(I, Q, U and S) from the cut sky maps. Since the ${\bf N}^{-1}$ matrices were calculated using the cut sky coverage, cut sky maps
with the appropriate mean value for use with these matrices may be obtained by simply masking the Galactic plane region.

\subsection{ Evaluation of the Inverse Pixel-Pixel Noise Matrix, ${\bf \Sigma}^{-1}$} \label{sec:inverse_pp_noise_matrix}
Several  steps in the \wmap~data processing and spectral processing require an explicit numerical representation of the
inverse pixel-pixel noise correlation matrix, ${\bf \Sigma}^{-1}$. Formally this matrix is described as
\begin{equation}
  {\bf \Sigma}^{-1} = {\bf M}^T {\bf N}^{-1} {\bf M } = \Sigma^{-1}(p_1, p_2) =  \sum_{t_1, t_2} M(t_1, p_1) N_{\rm tt}^{-1}(t_1 - t_2)M(t_2, p_2),
\end{equation}
where ${\bf N}^{-1}$ and ${\bf M}$ are the inverse noise matrix and the mapping matrix as described in 
\S~\ref{sec:map_evaluation} and \S~\ref{sec:radiometer_non-idealities}, and $p_1$ and $p_2$ are pixel indices spanning 
the I, Q, U  and S maps. The sums over $t_1$ and $t_2$ extend over all non-zero values of the 
inverse noise matrix evaluated at time $t_1-t_2$. These matrices were evaluated at  r4 on a year-by year basis
for all 10 DAs, resulting in 30 matrices. One approximation was used in the evaluation of these matrices to speed processing. 
Recall that each observation populates two rows of the mapping matrix, each row corresponding 
 the projection factors for each  of the two radiometers comprising the DA.
 Within each row the 8 non-zero elements correspond to
the two differential pixels (A-side and B-side) of the 4 maps I, Q, U and  S.
The approximation consists of grouping time contiguous observations for which both the A-side beam and B-side beam remain in the same
 r4 pixels. Each such group comprises $\approx 30$ rows of the mapping matrix (about 15 observations) whose non-zero elements reside
in the same 8 columns. For each  contiguous group of observations, the averages of each of the columns containing non-zero values 
is calculated for the rows associated with each radiometer. These averages are
tagged with the group's starting and ending times,  the pixel indices of the A-side and B-side beams and the radiometer identification.
 Given the starting and stopping 
times it is possible to calculate the appropriately weighted sum of $N_{\rm tt}^{-1}(t_1 - t_2)$ corresponding to  any pair of groups 
and corresponding radiometer. 
The sums over $t_1$ and $t_2$ are then calculated for all pairs of groups for which the weighted sum of $N_{\rm tt}^{-1}(t_1 - t_2)$ is non-zero.
This is accomplished  by multiplying this weighted sum by the average values of the appropriate columns of the 
corresponding groups, and accumulating the values into pixel-pixel noise matrix elements based on the pixel indices 
associated with each group. The result is a $4 N_{\rm p} \times 4 N_{\rm p}$ matrix, ${\bf \Sigma}^{-1}$, which may be inverted 
by standard numerical techniques to form the noise matrix ${\bf \Sigma}$. Using the group average values for the data in the columns 
of ${\bf M}$ rather than the exact row by row values in calculation of the inverse noise matrix 
is a good approximation for the vast majority of the sky,
since the values being averaged are smooth and slowly varying. However these assumptions fail for pixels very close to the Galactic poles 
since the polarizations angles $\gamma_{\rm A}$ and $\gamma_{\rm B}$ vary significantly \emph{within} a  r4 pixel. The effect of these
approximations has been investigated and is found to have negligible affect on the calculated power spectra. The r4 noise matrices
contained in the data release are calculated from mapping matrices corresponding to cut sky 
coverage and are intended for use in regions outside of the
region excluded by the processing mask. Values corresponding to r4 pixels entirely contained within 
the processing mask are set to zero.

\subsubsection{Projecting Transmission Imbalance Modes from the ${\bf \Sigma}^{-1}$ Matrices}
\begin{figure*}
\epsscale{1.0}
\plotone{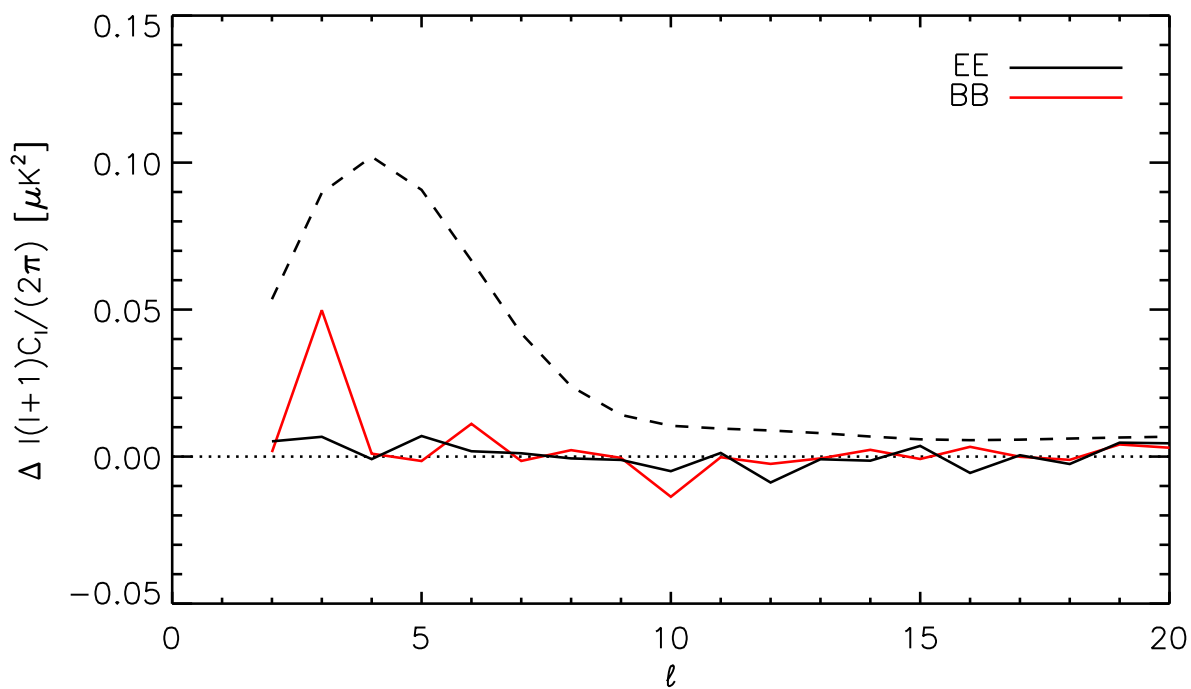}
\caption{Difference between the low-$\ell$ polarization power spectra with and without 
removal of the transmission-imbalance modes from the ${\bf N}^{-1}$ matrices. The black and red lines are the differences between 
the power spectra of combined Q-band and V-band polarization data for the E and B modes respectively. For comparison, the dashed line is the expected
E-mode signal corresponding to the best fit \wmap~power spectrum. The effect is largest for the B-mode $\ell = 3$ which 
already has a large uncertainty  arising from the scan pattern } 
\label{fig:loss_imbalance}
\end{figure*}

One potential source of systematic artifacts in the sky maps arises from the 
error in the determination of the input transmission imbalance
parameters, $x_{\rm im}$. These parameters are measured from the flight data and are used to 
calculate the estimated dipole signal that is removed from the TOD in preparation for 
the conjugate gradient map processing. Due to the large amplitude of the dipole signal, even small
errors in the measured values of transmission imbalance parameters could introduce significant artifacts
 into the sky maps. This effect was studied through the use of simulations in which a simulated TOD archive was produced
with a given set of $x_{\rm im}$ values and was analyzed with the input value of $x_{\rm im}$ as well as values $20\%$ 
higher and lower to simulate errors in the measurement of $x_{\rm im}$. Differences between the maps produced using the 
correct and incorrect $x_{\rm im}$ displayed well defined spatial structure confined to low $\ell$ with small variations
between DAs arising from the slightly different scan patterns. The geometry of these structures is completely determined by 
the scan patterns, while their amplitude in the final sky maps is determined by  the difference between the $x_{\rm im}$ value 
used to process the TOD 
and the ``true'' value of $x_{\rm im}$. The patterns from these simulation are therefore treated as templates, used to 
exclude the modes corresponding to the aforementioned spatial structures from subsequent analysis. This is accomplished through the use of 
modified versions of the inverse noise matrices, 
$\widetilde{{\bf \Sigma}}^{-1}$, 
that have these modes projected out. Each modified matrix is calculated as  
\begin{equation}
\widetilde{{\bf \Sigma}}^{-1} = {\bf \Sigma}^{-1} - \frac{{\bf \Sigma}^{-1}{\bf v}\otimes{\bf \Sigma}^{-1}{\bf v}}{{\bf v}^{\rm T}{\bf \Sigma}^{-1}{\bf v}}
\end{equation}
so that
\begin{equation}
\widetilde{{\bf \Sigma}}^{-1} {\bf v} = 0.
\end{equation}
In the above expressions  $\otimes$ designates an 
outer-product and ${\bf v }$ is the map mode to be removed, obtained from differences between the simulated 
maps made with the correct transmission imbalance factors and the incorrect factors. 
Using these forms of the matrices in the spectral analysis
\citep{page/etal:prep}  eliminates artifacts which might result from the 
errant signals cause by errors in the measured values of the transmission imbalance
parameters. Figure~\ref{fig:loss_imbalance} shows the effect of suppressing these modes on 
the low-$\ell$ polarization power spectra. This correction has a very small effect
on the cosmological parameters, but has been included in the results presented in ~\citet{page/etal:prep} for completeness.

Inverse noise \emph{covariance} matrices based on these \emph{correlation} 
matrices are contained in the data release. The covariance matrices are the correlation matrices 
scaled by a multiplicative factor, $1/\sigma_0^2$, where $\sigma_0$ is the noise per observation corresponding to each DA.

\subsection{ Production of Low Resolution Maps}
Evaluation of the low-$\ell$ power spectra with optimal signal-to-noise involves use of sky maps and the inverse 
pixel-pixel noise matrices, ${\bf \Sigma}^{-1}$, which describe the maps' noise properties. Since the matrices are produced
at  r4 due to computational constraints, corresponding low resolution sky maps must be produced that  preserve the signal 
and noise properties as well as possible. There are several difficulties involved with production of low resolution maps.
Simply binning maps to lower resolution, using either uniform or pixel-by-pixel (diagonal) inverse noise weighting will result 
in aliasing of high-$\ell$ signal and noise into lower-$\ell$ modes. This effect can be reduced by applying a low-pass 
filter to the map before degradation, but such a filter introduces additional noise correlations, not described 
by the inverse noise matrix as described in the previous section. Even without application of a filter, the omission of the
inter-pixel noise correlations results in the degraded maps having noise properties not precisely described by the noise matrix
as calculated in \S~\ref{sec:inverse_pp_noise_matrix}.
 
It is  possible to produce low resolution sky maps directly using the procedures described in 
\S~\ref{sec:map_evaluation} by accumulating data directly into low resolution maps. 
This corresponds to reducing the number of columns in the mapping matrix, ${\bf M}$, to reflect the reduced number of pixels
in the maps. Such maps will have noise properties accurately described by the pixel-pixel noise matrix, but will have 
a \emph{higher} degree of signal aliasing than maps initially produced at high resolution and subsequently degraded. This occurs since  
in evaluation of the sky map (see eq.[\ref{eqn:map_soln}]) the low resolution form of the mapping matrix is applied to
the data multiple times, each time introducing aliased signal components, whereas degrading a map initially produced at high 
resolution only introduces the aliased components once. 

For the Stokes I maps, which are highly signal dominated, degraded res 9 maps are clearly
preferable, since the small errors in the noise description are of no consequence and these maps
contain lower levels of aliased signal. For the Stokes Q and U maps the signal to noise
ratio in the low-$\ell$ regime is on order unity. Low resolution maps of Stokes Q and U were produced both directly at  r4 and
by degrading r9 maps without application of an anti-alias filter. Analysis of both sets of maps produced very similar low-$\ell$
polarization results for both simulated and flight data. Based on these results it was determined that the noise properties of the 
degraded form of low resolution polarization maps was adequately described by the pixel noise matrices, so the degraded form is used for
all three Stokes parameters.

Although the inter-pixel noise correlations are ignored in the map degradation, the noise correlations between the
Stokes Q and U signals \emph{within} each high resolution pixel are included. Two different versions of reduced
resolution maps are produced and are described in the following sections.

\subsubsubsection{Production of Full Sky Reduced Resolution Maps}\label{sec:red_res_maps}
The full sky reduced resolution maps (r4) are comprised of two components. The first component, which excludes the Galactic plane region, is
obtained by degrading the r9 cut sky (spm) maps and has noise properties described by the inverse noise matrices
(\S\ref{sec:inverse_pp_noise_matrix}). The second component, used in the Galactic plane region, results from the degradation of the 
combined maps (\S\ref{sec:combined_maps}) and has noise that is \emph{not} described by the
inverse noise matrices. The noise in this portion of the maps is approximately described by the number of observations values, $N_{\rm obs}$,
contained in the sky map data products.

\begin{figure*}
\epsscale{1.0}
\plotone{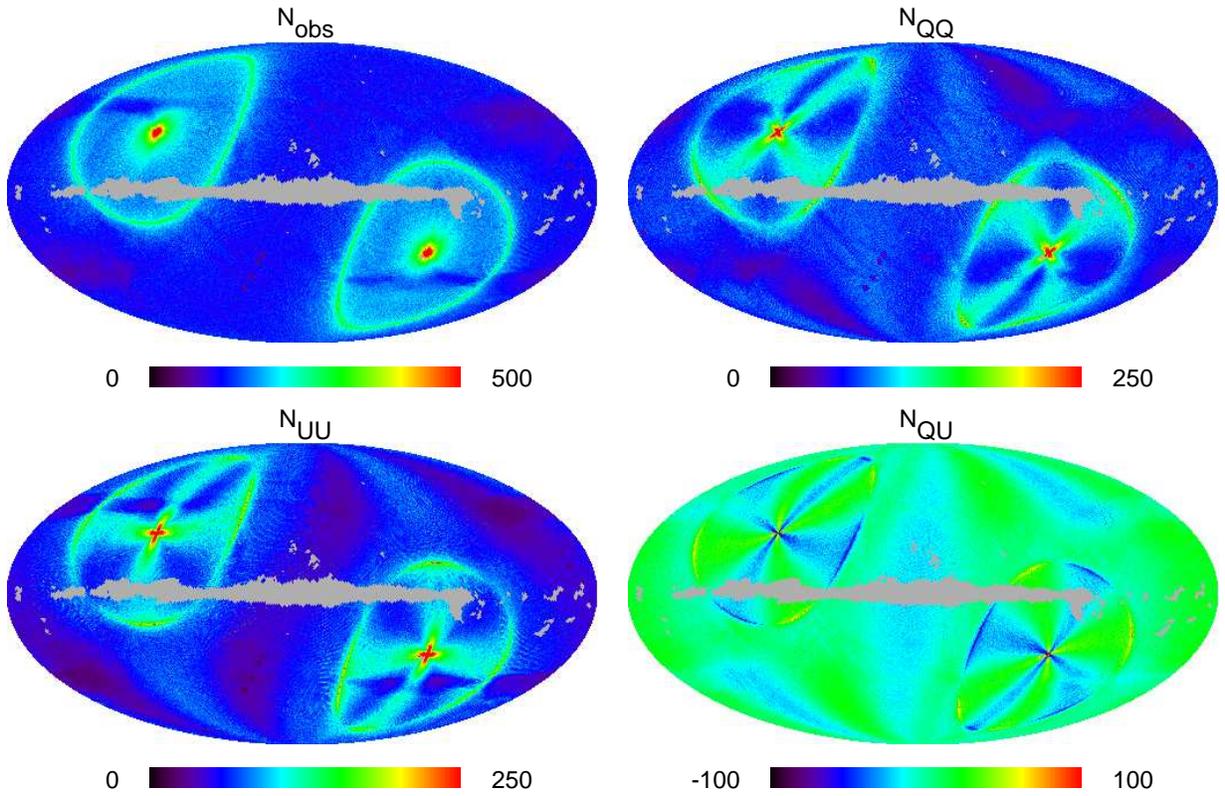}
\caption{Maps of $N_{\rm obs}$, $N_{\rm QQ}$, $N_{\rm UU}$, and $N_{\rm QU}$ weights for K band year-1 spm sky coverage. These maps
are measures of the noise covariance and are used when the maps are degraded from r9 to r4.} 
\label{fig:pass2_nobs_maps}
\end{figure*}

The first component of the reduced resolution map, comprising the high Galactic latitude regions, is generated as follows:
During production of the high resolution spm maps, the total weights applied to each pixel in the 
I, Q, U and S maps, $N_{\rm obs}$, $N_{\rm QQ}$, $N_{\rm UU}$, and $N_{\rm SS}$,
are calculated. These are simply the sums of the squares of the corresponding columns of the mapping
matrix, ${\bf M}$. The QU, QS and SU off-diagonal terms, $N_{\rm QU}$, $N_{\rm QS}$ and $N_{\rm SU}$, with each r9 pixel are also calculated.
These are the sums of the products of the corresponding columns of the mapping matrix.
 Figure~\ref{fig:pass2_nobs_maps} shows maps of the $N_{\rm obs}$, $N_{\rm QQ}$, $N_{\rm UU}$, and $N_{\rm QU}$ patterns.

For each high resolution pixel, $\rm p9$, an ${\bf N}_{\rm obs}$ matrix is formed, 
\begin{equation}
{\bf N}_{\rm obs}({\rm p9}) = \left( \begin{array}{ccc}
    N_{\rm QQ} & N_{\rm QU} & N_{\rm QS}\\ 
    N_{\rm QU} & N_{\rm UU} & N_{\rm SU}\\
    N_{\rm QS} & N_{\rm SU} & N_{\rm SS}
\end{array} \right ), 
\end{equation}
which differs only by a multiplicative constant from the inverse noise matrix. Correlations between the Stokes I and the polarization
related components are not used in the map degradation.
The Q, U and S values in a low resolution map pixel, $\rm p4$,  are then simply the inverse noise weighted average of all the Q, U and S values
of the high resolution pixels it contains,
\begin{eqnarray}
\left (   \begin{array}{c}
  Q_{\rm p4} \\
  U_{\rm p4} \\
  S_{\rm p4} 
\end{array} \right )& = & \left( {\bf N}_{\rm obs}^{\rm tot}({\rm p4}) \right)^{-1} 
\sum_{{\rm p9} \in {\rm p4}} w({\rm p9})~{\bf N}_{\rm obs}({\rm p9}) \left (   \begin{array}{c}
  Q_{\rm p9} \\
  U_{\rm p9} \\
  S_{\rm p9}
\end{array} \right ),\\
{\bf N}_{\rm obs}^{\rm tot}({\rm p4})& = &\sum_{{\rm p9} \in p4} {w({\rm p9})~\bf N}_{\rm obs}({\rm p9}).
\end{eqnarray}
Here $w({\rm p9})$ is the r9 processing mask~\citep{hinshaw/etal:prep} containing values of $0$ and $1$, and the summations
are for all r9 pixels contained within each r4 pixel. Elements of the r4 maps containing one or more observations are
retained. The Stokes I maps were degraded using simple $N_{\rm obs}$ weights.

Elements of the r4 maps described above with no observations are filled with values from degraded
forms of the combined maps (\S\ref{sec:combined_maps}). The combined maps are degraded following the same procedure, but
the ${\bf N}_{\rm obs}({\rm p9})$ matrices are formed 
using weight values from the cut sky (spm) maps when outside the processing mask, and values from the full sky (fs) maps
within the processing mask. 
Since the weights corresponding to the full sky
maps differ from those of the spm maps, the inverse noise matrices, calculated from spm map weights, do not describe the
noise in this portion of the map. All the rows and columns of the ${\bf \Sigma}^{-1}$ matrices corresponding to these 
pixels contain zeros.

\subsubsubsection{Production of the Partial Sky, Foreground Cleaned Reduced Resolution Polarization Maps}
The reduced resolution sky maps used in the pixel based low-$\ell$ likelihood evaluation~\citep{page/etal:prep} are formed
from the foreground cleaned  Q and U r9 spm maps, as described in~\citet{hinshaw/etal:prep} and~\citet{ page/etal:prep}.
Including the spurious term in the map degradation and likelihood evaluation was found to have a negligible effect on the
likelihood results~\citep{page/etal:prep}. Therefore, to simplify and speed calculation, only the Q and U terms
were used in the the degradation and low-$\ell$ likelihood evaluation of the foreground cleaned maps~\citep{page/etal:prep}.

The foreground cleaned Q and U maps were degraded in a similar fashion to the full sky maps described above: 
For each high resolution pixel, $\rm p9$, an ${\bf N}_{\rm obs}$ matrix is formed, 

\begin{equation}
{\bf N}_{\rm obs}({\rm p9}) = \left( \begin{array}{cc}
    N_{\rm QQ} & N_{\rm QU} \\ 
    N_{\rm UQ} & N_{\rm UU} \\ 
\end{array} \right ).
\end{equation}
 
The values of  the  low resolution map pixels, $\rm p4$,  are the inverse noise weighted sum of the values of all the constituent high resolution pixels,
\begin{eqnarray}
\left (   \begin{array}{c}
  Q_{\rm p4} \\
  U_{\rm p4} \\
\end{array} \right )& = & W({\rm p4}) \left( {\bf N}_{\rm obs}^{\rm tot}({\rm p4}) \right)^{-1} 
\sum_{{\rm p9} \in {\rm p4}} w({\rm p9}){\bf N}_{\rm obs}({\rm p9}) \left (   \begin{array}{c}
  Q_{\rm p9} \\
  U_{\rm p9} \\
\end{array} \right ),\\
{\bf N}_{\rm obs}^{\rm tot}({\rm p4})& = &\sum_{{\rm p9} \in {\rm p4}} {w({\rm p9}) \bf N}_{\rm obs}({\rm p9}).
\end{eqnarray}
In this case $w({\rm p9})$ is the r9 $P06$ mask~\citep{page/etal:prep} containing values of $0$ and $1$ and $W({\rm p4})$ is a low resolution 
form of this mask which contains zeros in elements for which more than half of the r9 pixels were masked and values of unity
otherwise.   
These degraded maps are contained in  the data release.

\section{TESTS ON THE DATA SET}\label{sec:tests_on_data}
\subsection{Limits on Spin-Synchronous Artifacts in the TOD}
Among the most troublesome type of systematic errors in experiments such as \wmap~are spurious signals which are synchronous with
the scanning motion of the instrument. Such artifacts do not integrate to lower levels with the inclusion of additional data
as does random noise. The most likely source of such driving signals is from  varying insolation due to the observatory's motion
driving temperature variations in the
instrument hardware, voltage fluctuation on the spacecraft power bus, or coupling directly into the microwave optics. For all of these
effects the errant signal is expected to be synchronous with the 129 s motion of the sun relative to the spacecraft. Following the 
procedure used in the year-1 analysis~\citep{jarosik/etal:2003b} the radiometer output signals were binned in a spacecraft fixed 
coordinate system based on the azimuth of the Sun about the spin axis of the spacecraft. When binned in such a manner any signal
synchronous with the azimuthal position of the sun about \wmap~should be preserved while asynchronous signal components should average away.
Combinations of radiometer outputs corresponding to both temperature anisotropies, $d1 + d2$, and polarization signals, $d1 - d2$, 
were accumulated for the entire 3 year mission. The resultant signals were all found to be less that $1.25~\ukelvinrms$. We therefore 
believe that there is little contamination of the TOD arising from spin synchronous solar effects and no corrections are applied
to the TOD for spin synchronous signals.

\subsection{Year-to-Year Null Tests}
\begin{figure*}
\epsscale{0.9}
\plotone{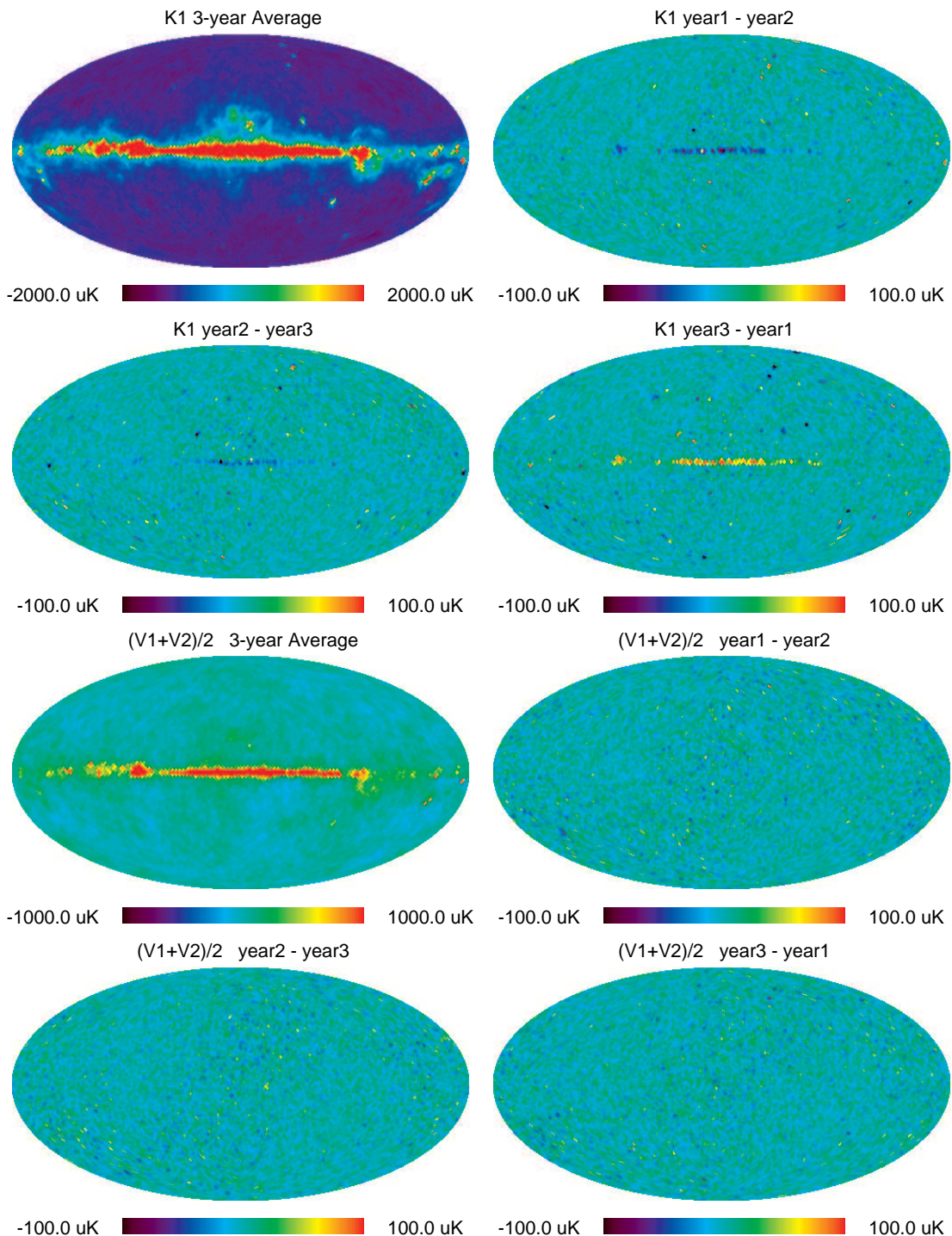}
\caption{\wmap~3-year combined and year-by-year difference maps of Stokes I for K-band and V-band. The residual
Galactic plane features in the K-band difference maps are consistent with the absolute calibration error of 0.5 \%. 
Several variable sources are also visible in the K-band maps. The CMB anisotropy signal is visible in the combined
V-band maps while no signal is visible in the V-band year-by-year difference maps. Maps are displayed at r5.} 
\label{fig:K1_V_I_diff_maps}
\end{figure*}

One of the most fundamental tests of the \wmap~maps is the year-to-year consistency between maps. Although such tests
do not eliminate the possibility of processing or systematic errors that repeat year to year, they do test for 
spurious signals such as striping from radiometer $1/f$ noise and glitches. A comparison between the year-1 and 3-year
processing of the first year data from the V2 DA was presented in \S~\ref{sec:gain_model_quad} and displayed a 
predominantly quadrupolar feature associated with the use of the improved radiometer gain model. Since both
maps used in that comparison  were generated from the same TOD, the white noise component largely canceled in the difference 
maps allowing this feature to be seen. In the case of year-to-year comparisons such white noise cancellations do not occur,
so subtle effects may not be evident. Nevertheless two sets of year-by-year difference maps are presented to demonstrate
the year-to-year consistency, one based on K bands with the largest foreground signal, and one based on V-band with the 
smallest foreground signal. The top four panels in Figure~\ref{fig:K1_V_I_diff_maps} display the 3 year average and year by year
differences of the K-band full sky maps. The 3-year combined map is dominated by Galactic emission and also displays a number
of point sources at high Galactic latitudes. The high level of Galactic emission makes K-band one of the more difficult
bands to calibrate. While the year-by-year null maps are noise dominated some residual Galactic plane 
signal is evident. This signal arises from small calibration errors of $\approx 0.1-0.2 \%$ and is consistent with the stated
calibration uncertainty of $0.5 \%$. Also evident in these maps are some features of both signs at high latitude arising from
known variable sources.

The lower four panels in Figure~\ref{fig:K1_V_I_diff_maps} display the 3-year average and year-by-year difference of the
V-band data. In this case Galactic emission can be seen at low latitudes and CMB anisotropies at high latitude. The difference
maps are again noise dominated with the spatially varying noise properties evident. The increased noise near the ecliptic plane
is a consequence  of the lower number of observations in these regions as determined by the scan strategy.

\begin{figure*}
\epsscale{0.9}
\plotone{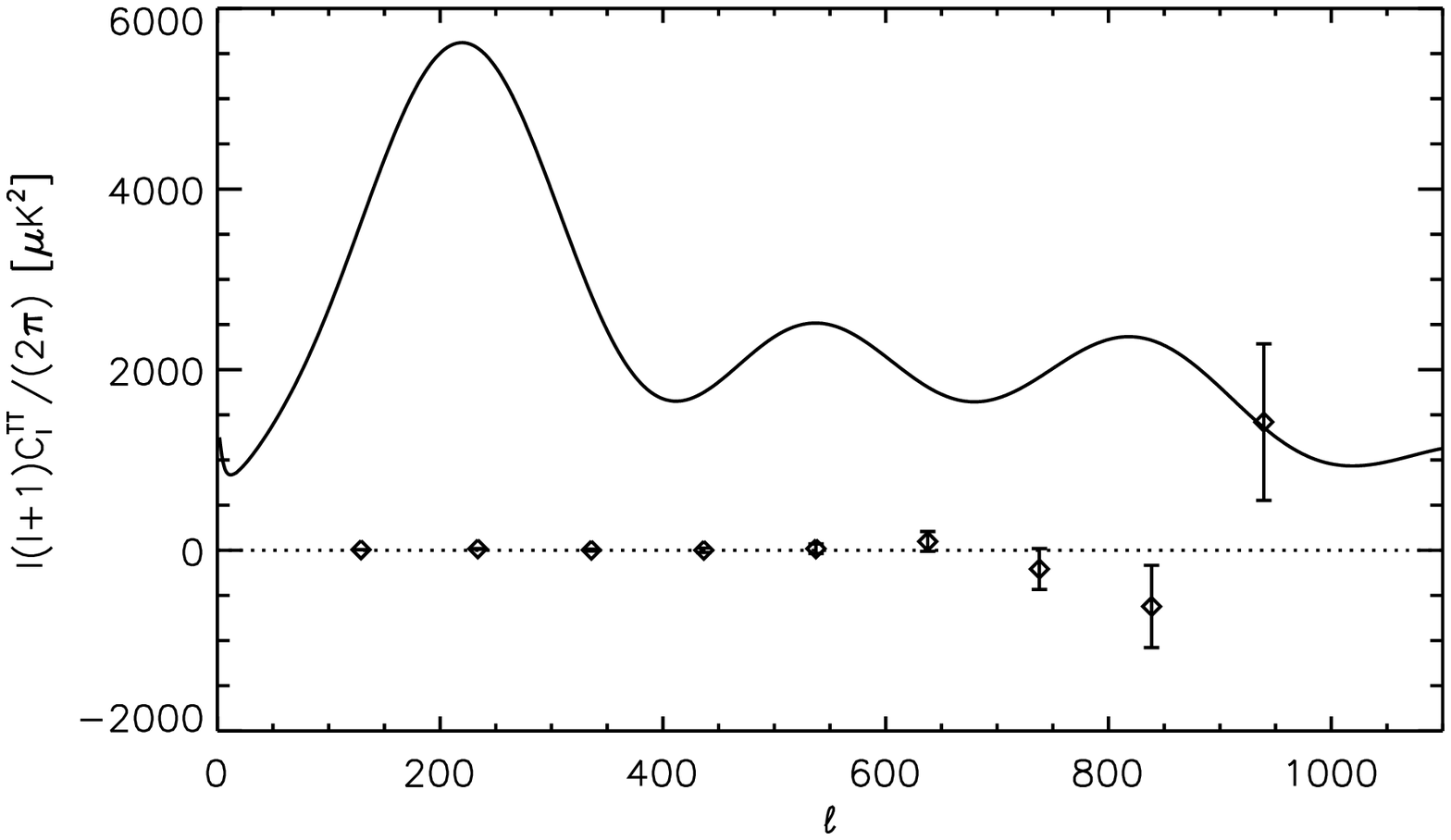}
\caption{Cross power spectrum of year-1 and year-2 null maps. For each year Stokes I null maps were formed 
for DA combinations ((V1-V2) + (W1-W2) + (W3-W4))/6, 
and the year-to-year cross spectrum calculated. Within each year the map differences remove true sky signals, and the year-to-year 
cross spectrum eliminates the noise bias. The power spectrum (diamonds) has been binned in $\Delta \ell =100$ to reduce noise. For comparison the
solid line is the best fit \wmap~Stokes I power spectrum. The null map residuals are very small relative to the measured power spectrum, reflecting 
the repeatability of the measurements } 
\label{fig:null_ps}
\end{figure*}

Simple power spectra of individual year-to-year null maps are dominated by the radiometer noise bias which grows as $\ell(\ell+1)$ when 
plotted with the customary scaling, and are not useful for assessing year-to-year map consistency. Figure~\ref{fig:null_ps} compares the 
best fit \wmap~CMB power spectrum with the year-1 to year-2 {\emph cross} power spectrum~\citep{hinshaw/etal:2003} of combined V and W band Stokes I sky maps,
averaged in $\Delta \ell$ bins of 100.  For each year, the linear combination of sky maps (V1-V2) + (W1-W2) + (W3-W4) was formed.
This combination removes the true sky signals to the extent that the bandpasses and gains of each pair of 
maps  are matched, while the use of the year-to-year cross spectrum eliminates the noise bias since the 
radiometer noise is uncorrelated year-to-year. The power spectrum of the null maps are very small relative to the measured power spectrum.

\begin{figure*}
\epsscale{0.9}
\plotone{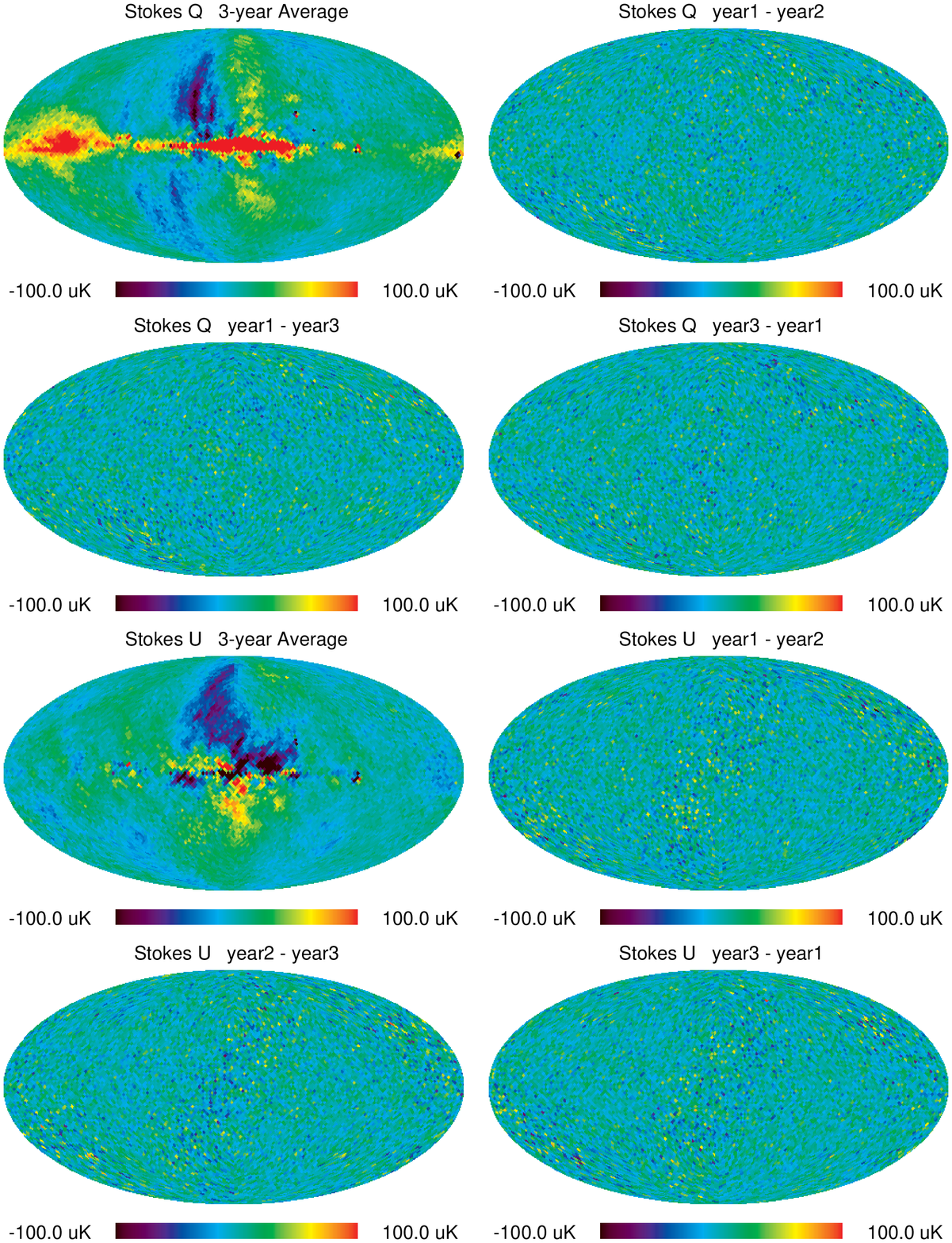}
\caption{\wmap~3-year combined and year-by-year difference maps of Stokes Q (top 4 panels) and U (bottom 4 panels) for K-band. Galactic
foreground emission is clearly visible in both Stokes Q and U of the three year combined maps. The difference
maps show no residual signal, although the spatially varying noise levels are apparent. Maps are displayed at r5. } 
\label{fig:V_QU_diff_maps}
\end{figure*}

A set of null maps for the Stokes Q and U parameters is displayed in Figure~\ref{fig:V_QU_diff_maps} for K-band.
Here in the year-combined maps the Galactic foreground emission, predominantly synchrotron, is clearly
evident. The year-by-year difference maps are again noise dominated. The spatially varying noise structure
for the Stokes Q and U maps is more complicated than that of the Stokes I maps due to the additional polarization
projection factors (e.g. $ \sin 2\gamma $) and the noise covariance between the Stoke Q and Stokes U maps of
each DA. The faint cross-like features near the ecliptic poles are regions with large weights, and therefore low 
statistical noise, as can be seen by comparing these null maps with the $N_{\rm obs}$ maps in Figure~\ref{fig:pass2_nobs_maps}.
Simple null tests, as displayed in Figure~\ref{fig:null_ps} for Stokes I, are not useful in assessing year-to-year consistency of the 
polarization maps due to the much smaller level of the expected signals and more complicated noise properties of the polarization maps.
Detailed analysis of the polarization results, including null tests, is presented in \citet{page/etal:prep}.

\subsection{Determination of Map Noise Levels}
\label{sec:map_noise}
Evaluation of the instrument noise level is an important aspect of 
many analyses involving the \WMAP~sky maps. In principle, a single noise per observation parameter, $\sigma_0$, 
should be sufficient to characterize the noise level
for each set of maps (a set consisting of the I, Q and U maps for a given year of data and DA). Such a characterization 
would require the use of full resolution pixel-pixel noise correlation matrices, 
describing the noise correlations between all combinations of pixel indices and Stokes components. 
For r9 sky maps, the size of such matrices precludes their use. However, it is possible to characterize the map noise levels
using only the diagonal components of the noise correlation matrices, $N_{\rm obs}$, $N_{\rm QQ}$, and $N_{\rm UU}$, provided the
\emph{intra-pixel} components of noise correlation matrix, $N_{\rm QU}$, $N_{\rm QS}$, and $N_{\rm SU}$ (\S~\ref{sec:red_res_maps}) are included.
This approximation uses different values of the noise per observation parameter for maps of the different Stokes components, parameterizing the 
effects of the ignored, off diagonal components of the noise correlation matrices. The different values of these 
parameters for the different map components result from the differences between the frequency distributions 
of the intensity and polarization information in the TOD, and the radiometers'
frequency dependent, i.e. 1/f, noise properties. The noise levels of the radiometers are also found to be stable year-to-year, so 
a single noise per observation value for each sky map/DA combination describes all three individual year sky maps. 

\subsubsection{Noise Evaluation of the r9 Maps}\label{sec:r9_noise}
The noise levels of the r9 maps in the current data release have been estimated using two different
methods. In the first method the noise-per-observation, is calculated directly for 
each DA from a weighted average of the variance
of year to year differences of sky maps. For the I maps the noise per observations based on year $x$ and year $y$ sky maps, $\sigma_0^{\rm xy, I}$,
 is given as
\begin{eqnarray}
\left(\sigma_0^{\rm xy, I}\right)^2 &=& N_{\rm pix}^{-1}\sum_p w(p) \frac{\left( t^{\rm x}(p) - t^{\rm y}(p)\right)^2}{n_{\rm obs}^{\rm x}(p)^{-1} +n_{\rm obs}^{\rm y}(p)^{-1}}\\
N_{\rm pix} &=& \sum_p w(p)
\end{eqnarray}
where $ t^{\rm x}(p)$ and $n_{\rm obs}^{\rm x}(p)$ are the Stokes I value and the number 
of observations for pixel $p$ of the year $x$ sky maps, and $w(p)$ is the processing mask,
used to exclude regions of high Galactic emission from the analysis. Extending this technique to the polarization maps
requires a generalization to allow for the noise covariance between the Stokes Q and U maps and the spurious map, S.
This is implemented using the relations
\begin{eqnarray}
\left(\sigma_0^{\rm xy, P}\right)^2 &=& N_{\rm pix}^{-1}\sum_p w(p) \cdot ({\bf Z}^{\rm xy}(p))^T~\left(\mathcal{ N}^{\rm xy}(p)\right)^{-1}~{\bf Z}^{\rm xy}(p)\\ 
{\bf Z}^{\rm xy}(p) &=& \left (\begin{array}{c}
  Q_{\rm p}^{\rm x} - Q_{\rm p}^{\rm y} \\
  U_{\rm p}^{\rm x} - U_{\rm p}^{\rm y} \\
  S_{\rm p}^{\rm x} - S_{\rm p}^{\rm y} 
\end{array} \right )\\
\mathcal{N}^{\rm xy}(p) &=& {\bf N}_{\rm obs}^x(p)^{-1} + {\bf N}_{\rm obs}^y(p)^{-1} \label{eqn:QUSnoisemat}
\end{eqnarray}
where ${\bf Z}^{\rm xy}(p)$ is a vector comprised of the differences between the year $x$ and year $y$ values of the Q, U and S maps
at pixel $p$, and ${\bf N}_{\rm obs}^{\rm x}(p)$ is the corresponding $3 \times 3$ element  matrix describing the 
intra-pixel noise correlations between the Q, U and S maps for pixel $p$ and year $x$.
 Relative calibration errors
between the two maps and the year to year differences in the distributions of 
observation pointings \emph{within} pixels will allow some sky signal to leak 
into the difference maps, biasing the resultant noise estimates high. Values of $\sigma_0^{\rm I}$ and $\sigma_0^{\rm P}$ obtained by 
averaging the values of the three possible year-by-year difference combinations are presented in Table~\ref{tab:noise}. 

The second technique also measures the noise from the year to year difference of the sky maps.
Difference maps of each Stokes parameter, I, Q and U,  are formed from  year $x$ and $y$ sky maps for each DA.
The effective number of observations for each pixel of these difference maps are calculated. 
For the Stokes I the effective number of observation  is given as
\begin{equation}
n_{\rm eff}^{\rm xy, I}(p) = 1/\left((n_{\rm obs}^{\rm x}(p))^{-1} + (n_{\rm obs}^{\rm y}(p))^{-1}\right) 
\end{equation}
and for the Stokes Q and U components it is given as
\begin{equation}
n_{\rm eff}^{\rm xy, Q}(p) = 1/\mathcal{N}^{\rm xy}_{\rm qq}(p) 
\end{equation}
and
\begin{equation}
n_{\rm eff}^{\rm xy, U}(p) = 1/\mathcal{N}^{\rm xy}_{\rm uu}(p) 
\end{equation}
where $\mathcal{N}^{\rm xy}_{\rm qq}$ and $\mathcal{N}^{\rm xy}_{\rm uu}$ are the diagonal components of $\mathcal{N}^{\rm xy}(p)$
as defined in equation~(\ref{eqn:QUSnoisemat}). For each Stokes component the following procedure is applied:
1) Pixels are divided into groups, $G_i$,  based on their values of $n_{\rm eff}$,
2) The variance of the difference maps for the pixels in each group, $\textrm{Var}(G_{\rm i})$,  are calculated as well as the mean value of 
$N_{\rm eff}$, $\overline{n_{\rm eff}(G_{\rm i})}$,
3) These values are fit with two  parameters, $\sigma_0^{\rm xy}$ and $\sigma_{\rm r}^{\rm xy}$ in the form
\begin{equation}
\textrm{Var}(G_{\rm i}) = (\sigma_0^{\rm xy})^2 \cdot \overline{n_{\rm eff}(G_{\rm i})} +  (\sigma_{\rm r}^{\rm xy})^2
\end{equation}
where $\sigma_{\rm r}^{\rm xy}$ is a parameter that allows for the presence of a statistically isotropic
signal that does not scale with $\overline{n_{\rm eff}(G_{\rm i})}$ . Pixels in high Galactic emission 
regions, as determined by the processing mask,  are excluded from the analysis. 
Fits of all three year-by-year difference maps are 
performed and the resultant values averaged together, these results 
are also presented in Table~\ref{tab:noise} and agree within $0.3 \%$ with the values obtained 
using the first technique. Since there is no significant disagreement between the methods we
use the results of the first technique for analysis. By comparing the noise level obtained from the 
year 3  -  year 2 difference maps with those from the year 2 - year 1 difference maps we find that the
sensitivity of the radiometers has decreased at most by $0.1 \%$ over the 1 year period between the mean
epoch of the data sets. 
 
 \begin{deluxetable}{cccccccc} 
\tablecaption{Noise Levels for r9 Maps }
\tablecomments{The values presented in the first 5 columns of this table are estimates of the noise 
levels of the Stokes I, Q and U parameters in the 3-year high resolution maps for each DA.
Each value was calculated by two different methods as described in  \S~\ref{sec:map_noise} and the results are
presenting in units of $\mkelvin \cdot N_{\rm obs}^{1/2}$. The
two methods agree to within $0.3 \%$ for both the high resolution Stokes I and polarization maps. The values in the last column are the ratios of the 
low resolution (r4) maps noise estimates, obtained using the noise correlation matrices (\S\ref{sec:r4_noise}), to the high resolution noise of the I maps
(\S\ref{sec:r9_noise}). Differences in the noise estimates arise from  ignoring pixel-pixel noise correlations both in the map degradation and 
the noise estimation of the high resolution maps. }
\tablecolumns{8}
\tablehead{
\colhead{} & \colhead{Method} & \colhead{1}& \colhead{2}& \colhead{1} & \colhead{2}& \colhead{2} & \colhead{} \\
\colhead{DA} & \colhead{Signal} & \colhead{$\sigma_0^{\rm I}$}& \colhead{$\sigma_0^{\rm I}$}& 
\colhead{$\sigma_0^{\rm P}$} & \colhead{$\sigma_0^{\rm Q}$}& \colhead{$\sigma_0^{\rm U}$} & \colhead{$\sigma_0^{\rm r4}/\sigma_0^I$}
} 
\startdata
K1    && 1.439 & 1.439    & 1.455 & 1.453 & 1.453 & 1.057\\
Ka1   && 1.464 & 1.462    & 1.483 & 1.484 & 1.480 & 1.019\\
Q1    && 2.245 & 2.244    & 2.269 & 2.273 & 2.272 & 1.027\\
Q2    && 2.135 & 2.140    & 2.156 & 2.157 & 2.156 & 1.016\\
V1    && 3.304 & 3.311    & 3.330 & 3.333 & 3.333 & 1.023\\
V2    && 2.946 & 2.951    & 2.970 & 2.970 & 2.974 & 1.015\\
W1    && 5.883 & 5.896    & 5.918 & 5.935 & 5.930 & 1.023\\
W2    && 6.532 & 6.550    & 6.571 & 6.587 & 6.578 & 1.024\\
W3    && 6.885 & 6.885    & 6.925 & 6.918 & 6.931 & 1.020\\
W4    && 6.744 & 6.765    & 6.780 & 6.793 & 6.800 & 1.033\\
\enddata
\label{tab:noise}
\end{deluxetable}

\subsubsection{ Noise Evaluation the r4 Full Sky Maps } \label{sec:r4_noise}
The noise of the full sky reduced resolution (r4) maps may be determined using the pixel-pixel noise matrices. This analysis
is performed only on the high Galactic latitude regions, those pixels originating
 from degradation of the high resolution spm maps. For each DA, noise levels are evaluated for all three possible
year-to-year difference map combinations. For combinations of years $\rm x$ and $\rm y$ the noise per 
observation, $\sigma_0^{\rm x,y}$, is calculated as
\begin{eqnarray}
(\sigma_0^{\rm x, y})^2 &=& ({\bf t}_{\rm x} - {\bf t}_{\rm y})^T {\bf \Sigma}_{\rm x,y}^{-1} ({\bf t}_{\rm x} - {\bf t}_{\rm y}),\\
{\bf \Sigma}_{\rm x, y}^{-1} &=& ({\bf \Sigma}_{\rm x} + {\bf \Sigma}_{\rm y})^{-1},
\end{eqnarray}
where ${\bf t}_x$ and ${\bf t}_y$ are reduced resolution sky map sets comprising I, Q, U and S components for
years x and y respectively, and ${\bf \Sigma}_{\rm x}$ and ${\bf \Sigma}_{\rm y}$ are the 
pixel-pixel noise correlation matrices for the corresponding years. For each DA, the values from the three 
year-to-year combinations are averaged together. The ratio of these average values to the noise values of 
the r9 I maps are presented in Table~\ref{tab:noise}. 
The differences between the two noise estimates arise from ignoring the pixel-pixel noise
correlations in both the r9 - r4 map degradation, and the evaluation of the $\sigma_0$ of the r9 maps. 

\section{SUMMARY}
\wmap~continued to operate normally throughout its second and third years of science observations. The additional observations
served not only to reduce the statistical noise, but also provided for improved understanding of the beam geometries
and radiometer performance, important factors in interpretation and processing of the science data. Updated versions  
of the radiometer gain model, beam models and window functions were presented. New processing procedures that
produce maximum likelihood sky maps for Stokes I, Q and U parameters have been described in detail as well as the calculation
of the inverse noise matrix. The techniques used to 
quantify  systematic errors were described along with details of the sky map processing used to correct for
potential systematic artifacts. Sky maps produced from data from different epochs were used to display the year to year
reproducibility of the signal and to measure the instrument noise levels. The performance of the \wmap~instrument remains
nominal with no significant decrease in sensitivity observed to date.

\acknowledgments
The \wmap~mission is a collaborative effort between Goddard Space Flight Center, Princeton University, the National 
Radio Astronomy Observatory and is supported by the Office of Space Sciences at NASA Headquarters. EK acknowledges 
support from the Alfred P. Sloan Fellowship.
 Some of the results in this paper have been derived using the HEALPix (G{\'o}rski, Hivon, and Wandelt 1999) package.

\appendix
\section{Sidelobe Correction and Rescaling }\label{app:sidelobe_corr}
Following a notation similar to that of \citet{hinshaw/etal:2003b} the \emph{uncalibrated} time ordered 
differential data for each detector, ${\bf c}$, may be represented as
\begin{equation}
{\bf c}  = {\bf g} \int d\Omega_{\bf n}{\bf M}({\bf n}){\bf t}({\bf n}). 
\end{equation}
Here ${\bf g}$ is instrument responsivity ($\rm cts/mK$), ${\bf M}({\bf n})$ is the mapping function
which transforms spatial information into temporal information and ${\bf t}({\bf n})$
is the sky signal being observed ($\rm mk$). All the aforementioned quantities are time dependent
as indicated by the use of boldface symbols.

The mapping function can be expressed in terms of the 
differential beam response of the instrument in instrument coordinates, $B_{\rm d}({\bf n}_{\rm inst})$, and 
${\bf R}$, the time dependent rotation
matrix that transforms sky coordinates, $n$, into the satellite coordinates, $n_{\rm inst}$.
\begin{equation}
{\bf M}({\bf n}) = B_{\rm d}({\bf R} \cdot {\bf n})
\end{equation}
The differential beam response is a signed quantity and is normalized such that
\begin{equation}
\int_{4\pi} d\Omega_{\bf n} |B_{\rm d}({\bf n})| = 2. 
\end{equation}
The fraction of the differential beam response in the sidelobes, $f_{\rm sl}$, is obtained by
integrating the magnitude of the  differential beam response over the region of the
map further than $r_{\rm H}$ from the boresights of both main beams,
\begin{equation}
f_{\rm sl} = \frac{1}{2} \int_{d_{\rm A}, d_{\rm B} > r_{\rm H}} d\Omega_{\bf n} |B_{\rm d}^{\rm sl}({\bf n})|,
\end{equation}
leaving the fraction $(1-f_{\rm sl})$ of the gain in the main beams.
As described in \citet{hinshaw/etal:2003b}, the calibration of \wmap
~is tied to the amplitude of the annual dipole signal arising from
the motion of \wmap ~with respect to the solar system barycenter. 
If the annual dipole sky signal is represented as ${\bf t}_{\rm ad}$, the corresponding 
component of the uncalibrated radiometer signal is
\begin{equation}
{\bf c}_{\rm ad}  =  {\bf g} \int d\Omega_{{\bf n}}{\bf M}({\bf n}){\bf t}_{\rm ad}({\bf n}). 
\end{equation}
Calibration is accomplished by performing a 
least-squares fit of the uncalibrated data to a predicted time ordered data set, ${\bf d}_{\rm ad}^{\rm mono}$, based
on the known monopole temperature, 
\begin{equation}
{\bf c}_{\rm ad} = {\bf\tilde{g}}~{\bf d}_{\rm ad}^{\rm mono}, 
\end{equation}
yielding an estimate of the radiometer gain, ${\bf \tilde{g}}$.
The predicted time ordered data set is obtained using a mapping function assuming ideal
`pencil' beams, ${\bf M}^{\rm pencil}$ and the annual dipole sky signal,
\begin{equation}
{\bf d}_{\rm ad}^{\rm mono}  =  \int d\Omega_{\bf n} {\bf M}^{\rm pencil}({\bf n}) {\bf \tilde{t}}_{\rm ad}^{\rm mono}({\bf n}).
\end{equation}
The actual calibration procedure is iterative and provides for subtraction of better estimates of the 
fixed sky signal before each new gain determination. Once the iterative steps are completed the radiometer gain
 model (Sec.~\ref{sec:gain_model}) is fit to smooth and interpolate the hourly ${\bf \tilde{g}}$ values.
 Taken together, the above calibration procedure effectively measures ${\bf \tilde{g}}$
by evaluating the mean of the ratio
\begin{equation}
{\bf \tilde{g}} =\left <\frac{{\bf g}  \int d\Omega_{\bf n} {\bf M}({\bf n}){\bf t}_{\rm ad}({\bf n})}{ \int d\Omega_{\bf n}
	{\bf M}_{\rm pencil}({\bf n}){\bf \tilde{t}}_{\rm ad}^{\rm mono}({\bf n})} \right >.
\end{equation}
Note that the mapping function in the numerator incorporates the full sky beam response. This function can be
written as the sum of a main beam mapping function, ${\bf M}^{\rm mb}$, and a sidelobe mapping function, ${\bf M}^{\rm sl}$, 
\begin{equation}
{\bf M} = {\bf M}^{\rm mb} + {\bf M}^{\rm sl}.
\end{equation}
In terms of these quantities the estimate of the radiometer gain becomes
\begin{equation}
{\bf \tilde{g}} =\left <\frac{{\bf g}  \int d\Omega_{\bf n} {\bf M}^{\rm mb}({\bf n}){\bf t}_{\rm ad}({\bf n}) + {\bf g} \int d\Omega_{\bf n} 
	{\bf M}^{\rm sl}({\bf n}){\bf t}_{\rm ad}({\bf n}) }{\int d\Omega_{\bf n} {\bf M}_{\rm pencil}({\bf n}){\bf \tilde{t}}_{\rm ad}^{\rm mono}({\bf n})}\right >.
\end{equation}
This can be approximated as
\begin{equation}
{\bf \tilde{g}} \approx {\bf g} (  1 - f_{\rm sl} + {\bf \epsilon}f_{\rm sl} )\frac{\tilde{T}^{\rm mono}}{T^{\rm mono}}. \label{eq:gain_approx}
\end{equation}
The response of the telescope main beam to a dipole signal is nearly identical to an ideal pencil beam except
that only the fraction $(1-f_{\rm sl})$ of the antenna gain is contained in the main beam, leading to the first two terms in the
parenthesis in equation~\ref{eq:gain_approx}. The last parenthesized term originates from the annual dipole signal
picked up through the sidelobe response, with ${\bf \epsilon}$ being a coupling efficiency factor parameterizing
how well the dipole signal couples to the sidelobe response pattern. The product ${\bf \epsilon}f_{\rm sl}$ was
evaluated by generating one year of simulated TOD representing the sidelobe response to a predicted 
annual dipole signal. This TOD was fitting in one hour periods
to a template function corresponding to the expected pencil beam response signal.
Although the quantity ${\bf \epsilon}$ is in principle time dependent, it was found to vary by less than 
10\% over the entire year simulation, and it is approximated as constant. Applying this calibration factor 
to the uncalibrated data yields a calibrated archive, 
\begin{equation}
{\bf d}  = \frac{\bf c}{\bf \tilde{g}} =   \frac{1}{1 - f_{\rm sl} + {\bf \epsilon} f_{\rm sl}} \int d\Omega_{\bf n}
	{\bf M}({\bf n}){\bf t}({\bf n}). \label{eq:cal_tod}
\end{equation}

A sidelobe corrected archive, calibrated to correspond to the main beam containing \emph{all} the antenna gain
can be written as
\begin{equation}
{\bf d}^{\rm mb}  =   \frac{1}{1 - f_{\rm sl}} \int d\Omega_{\bf n}{\bf M}^{\rm mb}({\bf n}){\bf t}({\bf n}), 
\end{equation}
where the prefactor of the integral compensates for the fact that only the fraction $(1-f_{\rm sl})$ of the beam response
is contained in the main beam in ${\bf M}^{\rm mb}({\bf n})$. To first order in $f_{\rm sl}$ this can be obtained from
the uncorrected calibrated archive, 
\begin{equation}
{\bf \tilde{d}}^{\rm mb}  \approx (1 + \epsilon f_{\rm sl}){\bf d} -  \int d\Omega_{\bf n}{\bf M}^{\rm sl}({\bf n}){\bf \tilde{t}}({\bf n}), 
\end{equation}
where ${\bf \tilde{t}}({\bf n}) = {\bf \tilde{t}_{\rm fixed} + {\bf \tilde{t}}_{\rm ad}^{\rm mono}}$ is an approximate sky map obtained 
from applying the iterative map making
algorithm to the calibrated (but not sidelobe corrected) TOD archive given by equation~\ref{eq:cal_tod}. Values of the recalibration factor, 
$(1 + \epsilon f_{\rm sl})$, are presented in Table~\ref{tab:intensity_processing1}.

\end{document}